\documentclass[a4paper, fleqn, usenatbib]{mnras}

\usepackage{newtxtext, newtxmath}

\usepackage[T1]{fontenc}
\usepackage{ae, aecompl}

\usepackage{graphicx}	
\usepackage{amsmath}	
\usepackage{amssymb}	
\usepackage{booktabs}
\usepackage{subcaption}
\usepackage{pdflscape}
\newcommand{\diff}{\mathrm{d}}

\DeclareMathOperator{\tr}{Tr}
\newcommand{\lensfit}{\emph{lens}fit}
\newcommand{\lcdm}{$\Lambda$CDM}

\title[The KiDS-450 shear power spectrum]{KiDS-450: The tomographic weak lensing power spectrum and constraints on cosmological parameters}

\author[F. K\"ohlinger et al.]{
F. K\"ohlinger,$^{1 \, ,2}$\thanks{E-mail: fabian.koehlinger@ipmu.jp}
M. Viola,$^{2}$
B. Joachimi,$^{3}$
H. Hoekstra,$^{2}$
E. van Uitert,$^{3}$
\newauthor{
H. Hildebrandt,$^{4}$
A. Choi,$^{5\, ,6}$
T. Erben,$^{4}$
C. Heymans,$^{5}$
S. Joudaki,$^{7\, ,8}$
D. Klaes,$^{4}$}
\newauthor{
K. Kuijken,$^{2}$
J. Merten,$^{9}$
L. Miller,$^{9}$
P. Schneider$^{4}$
and E.A. Valentijn$^{10}$}
\\
$^{1}$Kavli Institute for the Physics and Mathematics of the Universe (Kavli IPMU, WPI), The University of Tokyo Institutes for Advanced Study, \\
The University of Tokyo, Kashiwa, Chiba 277-8583, Japan\\
$^{2}$Leiden Observatory, Leiden University, PO Box 9513, Leiden, NL-2300 RA, the Netherlands\\
$^{3}$Department of Physics and Astronomy, University College London, Gower Street, London WC1E 6BT, UK\\
$^{4}$Argelander-Institut f\"ur Astronomie, Auf dem H\"ugel 71, 53121 Bonn, Germany\\
$^{5}$Scottish Universities Physics Alliance, Institute for Astronomy, University of Edinburgh, Royal Observatory,\\
Blackford Hill, Edinburgh, EH9 3HJ, UK\\
$^{6}$Center for Cosmology and AstroParticle Physics, The Ohio State University, 191 West Woodruff Avenue, Columbus, OH 43210, USA\\
$^{7}$Centre for Astrophysics \& Supercomputing, Swinburne University of Technology, PO Box 218, Hawthorn, VIC 3122, Australia\\
$^{8}$ARC Centre of Excellence for All-sky Astrophysics (CAASTRO)\\
$^{9}$Department of Physics, University of Oxford, Denys Wilkinson Building, Keble Road, Oxford OX1 3RH, UK\\
$^{10}$Kapteyn Institute, University of Groningen, PO Box 800, NL-9700 AV Groningen, the Netherlands\\}

\pubyear{2017}

\begin{document}
\label{firstpage}
\pagerange{\pageref{firstpage}--\pageref{lastpage}}
\maketitle

\begin{abstract}
We present measurements of the weak gravitational lensing shear power spectrum based on $450 \deg^2$ of imaging data from the Kilo Degree Survey. We employ a quadratic estimator in two and three redshift bins and extract band powers of redshift auto-correlation and cross-correlation spectra in the multipole range $76 \leq \ell \leq 1310$. The cosmological interpretation of the measured shear power spectra is performed in a Bayesian framework assuming a \lcdm \ model with spatially flat geometry, while accounting for small residual uncertainties in the shear calibration and redshift distributions as well as marginalising over intrinsic alignments, baryon feedback and an excess-noise power model. Moreover, massive neutrinos are included in the modelling. The cosmological main result is expressed in terms of the parameter combination  $S_8 \equiv \sigma_8 \sqrt{\Omega_{\rm m}/0.3}$ yielding $S_8 = \ 0.651 \pm 0.058$ (3 z-bins), confirming the recently reported tension in this parameter with constraints from \textit{Planck} at $3.2\sigma$ (3 z-bins). We cross-check the results of the 3 z-bin analysis with the weaker constraints from the 2 z-bin analysis and find them to be consistent. The high-level data products of this analysis, such as the band power measurements, covariance matrices, redshift distributions, and likelihood evaluation chains are available at \url{http://kids.strw.leidenuniv.nl}.
\end{abstract}

\begin{keywords}
cosmology: observations -- large-scale structure of Universe -- cosmological parameters -- gravitational lensing: weak.
\end{keywords}



\section{Introduction}
\label{sec:intro}

The current cosmological concordance model successfully describes observations spanning a wide range in cosmic volume from the cosmic microwave background (CMB) power spectrum (e.g. \citealt{Planck2015_CP}), the Hubble diagram based on supernovae of type IA (e.g. \citealt{Riess2016}), big bang nucleosynthesis (e.g. \citealt{FieldsOlive2006}), to the distance scales inferred from baryon acoustic oscillations imprinted in the large-scale clustering of galaxies (e.g. \citealt{BOSS2015}). Based on Einstein's theory of general relativity and the application of the Copernican principle to the whole Universe, the $\Lambda$-dominated cold dark matter ($\Lambda$CDM) model requires in its simplest form only a handful of parameters to fit all current observational data. 

The weak gravitational lensing due to all intervening cosmic large-scale structure along an observer's line-of-sight, termed cosmic shear, presents a powerful tool to study the spatial and temporal distribution of the dark species. However, the tiny coherent image distortions, the shear, of background sources caused by the differential deflection of light by foreground masses can only be studied in statistically large samples of sources. Hence, wide-field surveys covering increasingly more volume of the Universe provide the strategy for improving the precision of the measurements. 
Data from large weak lensing surveys such as the Kilo Degree Survey (KiDS; \citealt{KiDS2013, deJong2015, deJong2017}, \citealt{Kuijken2015}), the Subaru Hyper SuprimeCam lensing survey (HSC; \citealt{Miyazaki2015, HSC2017}), and the Dark Energy Survey (DES; \citealt{Jarvis2015}) are currently building up. 
These surveys are expected to reach a sky coverage on the order of (several) 1000 $\deg^2$ within the next few years, which presents an order of magnitude increase of data useful for cosmic shear studies compared to currently available survey data \citep{Erben2013, Moraes2014, RCS2016}. Eventually, close to all-sky surveys will be carried out over the next decade by the ground-based Large Synoptic Survey Telescope (LSST; \citealt{LSST2008}), or the spaceborne \textit{Euclid} satellite \citep{Euclid}. In contrast to that the spaceborne \textit{Wide Field Infrared Survey Telescope (WFIRST\footnote{\url{wfirst.gsfc.nasa.gov}}}) will only observe on the order of 1000 $\deg^2$ but to unprecedented depth.
%
The cosmic shear signal as a function of redshift is sensitive to the growth of structure and the geometry of the Universe and studying its redshift dependence allows us to infer the expansion rate as well as the clustering behaviour of cosmic species such as dark matter, massive neutrinos, and dark energy.

Several statistics have been used to measure cosmic shear; the most common one to date is based on the two-point statistics of real-space correlation functions (e.g. \citealt{Kilbinger2015} for a review). The redshift dependence is either considered by performing the cosmic shear measurement in tomographic redshift slices (e.g. \citealt{Benjamin2013}, \citealt{Heymans2013}, \citealt{Becker2015}) or by employing redshift-dependent spherical Bessel functions \citep{Kitching2014}.  
An alternative approach is to switch to Fourier-space and measure the power spectrum of cosmic shear instead. 
One particular advantage of direct shear power-spectrum estimators over correlation-function measurements is that the power-spectrum measurements are significantly less correlated on all scales. This is very important for the clean study of scale-dependent signatures, for example massive neutrinos, as well as to investigate residual systematics. For correlation functions accurate modelling is required for highly non-linear scales in order to avoid any bias in the cosmological parameters. Moreover, correlation-function measurements require a careful assessment and correction of any global additive shear bias. 

Direct power spectrum estimators have been applied to data a handful of times. The quadratic estimator \citep{Hu2001} was applied to the COMBO-17 dataset \citep{Brown2003} and the GEMS dataset \citep{Heymans2005}. In a more recent study, \citet{Lin2012} applied the quadratic estimator and a direct pseudo-$C(\ell)$ estimator \citep{Hikage2011} to data from the SDSS Stripe 82. 
However, the direct power spectrum estimators in these studies did not employ a tomographic approach. This was introduced for the first time in \citet{Koehlinger2016}, where we extended the quadratic estimator formalism to include redshift bins and applied it to shear catalogues from the lensing analysis of the Canada--France--Hawaii Telescope Legacy Survey (CFHTLenS; \citealt{Erben2013}, \citealt{Heymans2012}, \citealt{Hildebrandt2012}). 

For this paper we apply the quadratic estimator in two and three redshift bins to $450 \deg^2$ of imaging data from the Kilo Degree Survey (KiDS-450 in short hereafter). By comparing the results obtained here to results from the fiducial correlation-function analysis by \citet{Hildebrandt2016} we point out particular advantages and disadvantages of the quadratic estimator in comparison to correlation functions. Moreover, this analysis presents an important cross-check of the robustness of the cosmological constraints derived by \citet{Hildebrandt2016}, which were found to be in mild tension in the parameter combination $S_8 \equiv \sigma_8 \sqrt{\Omega_{\rm m}/0.3}$ at $2.3\sigma$ when compared to the most recent CMB constraints by \citet{Planck2015_CP}. 
The estimator and data extraction and cosmological inference pipelines used in this analysis are independent from the estimator and pipelines used in \citet{Hildebrandt2016}. Only the data input in the form of shear catalogues and redshift catalogues are shared between the two analyses. 

The paper is organized as follows: in Section~\ref{sec:theo} we summarise the theory for cosmic shear power spectra and in Section~\ref{sec:qe} we present the quadratic estimator algorithm. Section~\ref{sec:data_meas} introduces the KiDS-450 dataset, the applied shear calibrations, and the details of the employed covariance matrix of the shear power spectra. In Section~\ref{sec:signal} we present the measured cosmic shear power spectra. The results of their cosmological interpretation are discussed in Section~\ref{sec:results}. We summarise all results and conclude in Section~\ref{sec:conclusions}.

\section{Theory}
\label{sec:theo}

Gravitational lensing describes the deflection of light due to mass, following  from Einstein's principle of equivalence. In this paper we will specifically work in the framework of weak gravitational lensing. 
It is called weak lensing because the coherent distortions of the image shapes of galaxies are typically much smaller than their intrinsic ellipticities.
Measurements of the coherent image distortions are only possible in a statistical sense and requires averaging over large samples of galaxies due to the broad distribution of intrinsic ellipticities of galaxies.
The weak lensing effect of all intervening mass between an observer and all sources along the line-of-sight is called cosmic shear. The resulting correlations of galaxy shapes can be used to study the evolution of the large-scale structure and therefore cosmic shear has become an increasingly valuable tool for cosmology especially in the current era of large surveys (see \citealt{Kilbinger2015} for a review). For details on the theoretical foundations of (weak) gravitational lensing we refer the reader to the standard literature (e.g. \citealt{BartelmannSchneider2001}).    

The main observables in a weak lensing survey are the angular positions, shapes, and (photometric) redshifts of galaxies. The measured galaxy shapes in terms of ellipticity components $\epsilon_{1}$, $\epsilon_{2}$ at angular positions $\bmath{n}_i$ are binned into pixels $i = 1,\, ..., \, N_{\rm pix}$ and (photometric) redshift bins $z_\mu$. Averaging the ellipticities in each pixel yields estimates of the components of the spin-2 shear field, $\gamma(\bmath{n}, \, z_\mu) = \gamma_1(\bmath{n}, \, z_\mu) + {\rm i} \gamma_2(\bmath{n}, \, z_\mu)$. Its Fourier decomposition can be written in the flat-sky limit\footnote{This is well justified for the range of multipoles accessible with the current KiDS-450 data.} (see \citealt{Kilbinger2017}) as  
\begin{equation}
\begin{split}
\gamma_1(\bmath{n}, \, z_\mu)\pm \rm{i} \gamma_2(\bmath{n}, \, z_\mu) = \int & \frac{\diff^2 \ell}{(2 \upi)^2} W_{\rm pix}(\bmath{\ell}) \\
& \times [\kappa^{\rm E}(\bmath{\ell}, \, z_\mu) \pm \rm{i} \kappa^{\rm B}(\bmath{\ell}, \, z_\mu)] \\
& \times e^{\pm 2\rm{i} \varphi_\ell} e^{\rm{i} \bmath{\ell}\cdot\bmath{n}} \, ,
\end{split}
\end{equation}
with $\varphi_\ell$ denoting the angle between the two-dimensional vector $\bmath{\ell}$ and the $x$-axis. 

In the equation above we introduced the decomposition of the shear field into curl-free and divergence-free components, i.e. E- and B-modes, respectively. For lensing by density perturbations the convergence field $\kappa^{\rm E}$ contains all the cosmological information and the field $\kappa^{\rm B}$ usually vanishes in the absence of systematics. In the subsequent analysis we will still extract it and treat it as a check for residual systematics in the data. 

The Fourier transform of the pixel window function, $W_{\rm pix}(\bmath{\ell})$, can be written as
\begin{equation}
\label{eq:pixel_window}
W_{\rm pix}(\bmath{\ell}) = j_0 \left(\frac{\ell \sigma_{\mathrm{pix}}}{2} \cos \varphi_\ell \right) j_0 \left(\frac{\ell \sigma_{\mathrm{pix}}}{2} \sin \varphi_\ell \right) \, ,
\end{equation}
where $j_0(x)=\sin(x)/x$ is the zeroth-order spherical Bessel function and $\sigma_{\mathrm{pix}}$ is the side length of a square pixel in radians.

The shear correlations between pixels $\bmath{n}_i$ and $\bmath{n}_j$ and tomographic bins $\mu$ and $\nu$ can be expressed in terms of their power spectra and they define the shear-signal correlation matrix \citep{Hu2001}:
\begin{equation}
\label{eq:shear_corr_matrix}
\mathbfss{C}^{\mathrm{sig}} = \langle \gamma_a(\bmath{n}_i, \, z_\mu) \gamma_b(\bmath{n}_j, \, z_\nu) \rangle \, ,
\end{equation} 
with components
\begin{align} \label{eq:shear_corr}
\langle \gamma_{1i\mu} \gamma_{1j\nu} \rangle & = \int \frac{\diff^2 \ell}{(2 \upi)^2} [C^{\mathrm{EE}}_{\mu\nu}(\ell)\cos^2 2\varphi_\ell \nonumber \\
& \hphantom{{} = \int} + C^{\mathrm{BB}}_{\mu\nu}(\ell) \sin^2 2\varphi_\ell  \nonumber \\
& \hphantom{{} = \int} -C^{\mathrm{EB}}_{\mu\nu}(\ell) \sin 4  \varphi_\ell]W_{\rm pix}^2(\bmath{\ell})e^{{\rm i}\bmath{\ell}\cdot(\bmath{n}_i-\bmath{n}_j)} \, , \nonumber \\
\langle \gamma_{2i\mu} \gamma_{2j\nu} \rangle & = \int \frac{\diff^2 \ell}{(2 \upi)^2} [C^{\mathrm{EE}}_{\mu\nu}(\ell)\sin^2 2\varphi_\ell \nonumber \\
& \hphantom{{} = \int} +C^{\mathrm{BB}}_{\mu\nu}(\ell) \cos^2 2\varphi_\ell \nonumber \\ 
& \hphantom{{} = \int} +C^{\mathrm{EB}}_{\mu\nu}(\ell) \sin 4 \varphi_\ell]W_{\rm pix}^2(\bmath{\ell})e^{{\rm i}\bmath{\ell}\cdot(\bmath{n}_i-\bmath{n}_j)} \, , \nonumber \\
\langle \gamma_{1i\mu} \gamma_{2j\nu} \rangle & = \int \frac{\diff^2 \ell}{(2 \upi)^2} [\tfrac{1}{2} \, (C^{\mathrm{EE}}_{\mu\nu}(\ell) - C^{\mathrm{BB}}_{\mu\nu}(\ell)) \sin 4\varphi_\ell \nonumber \\ 
& \hphantom{{} = \int} +C^{\mathrm{EB}}_{\mu\nu}(\ell) \cos 4 \varphi_\ell ] W_{\rm pix}^2(\bmath{\ell})e^{{\rm i}\bmath{\ell}\cdot(\bmath{n}_i-\bmath{n}_j)} \, .
\end{align}

In the absence of systematic errors and shape noise\footnote{In lensing this term refers to a shot noise-like term that depends on the number of available source galaxies and their intrinsic ellipticity dispersion.}, the cosmological signal is contained in the E-modes and their power spectrum is equivalent to the convergence power spectrum, i.e. $C^{\mathrm{EE}}(\ell)=C^{\kappa\kappa}(\ell)$ and $C^{\mathrm{BB}}(\ell) = 0$. Shot noise will generate equal power in E- and B-modes. The cross-power between E- and B-modes, $C^{\mathrm{EB}}(\ell)$, is expected to be zero because of the parity invariance of the shear field. 

The theoretical prediction of the convergence power spectrum per redshift-bin correlation $\mu$, $\nu$ in the (extended) Limber approximation (\citealt{Limber1953}, \citealt{Kaiser1992}, \citealt{LoVerde2008}) can be written as: 
\begin{equation}
\label{eq:theo_power_spec}
C_{\mu \nu}^{\mathrm{EE}}(\ell) = \int_{0}^{\chi_{\rm H}} \diff \chi \, \frac{q_\mu(\chi) q_\nu(\chi)}{f_\mathrm{K}^2(\chi)} P_\delta\left(k=\frac{\ell + 0.5}{f_\mathrm{K}(\chi)}; \chi \right) \, ,
\end{equation}
which depends on the comoving radial distance $\chi$, the comoving distance to the horizon $\chi_{\rm H}$, the comoving angular diameter distance $f_\mathrm{K}(\chi)$, and the three-dimensional matter power spectrum $P_\delta(k; \chi)$. 

The weight functions $q_\mu(\chi)$ depend on the lensing kernels and hence they are a measure of the lensing efficiency in each tomographic bin $\mu$:
\begin{equation}
\label{eq:lensing_kernel}
q_\mu(\chi) = \frac{3\Omega_{\rm m}H_0^2}{2c^2} \frac{f_\mathrm{K}(\chi)}{a(\chi)}\int_\chi^{\chi_\mathrm{H}} \diff \chi^{\prime} \, n_\mu(\chi^{\prime})\frac{f_\mathrm{K}(\chi^{\prime}-\chi)}{f_\mathrm{K}(\chi^{\prime})} \, ,
\end{equation}
where $a(\chi)$ is the scale factor and the source redshift distribution is denoted as $n_\mu(\chi) \, \diff \chi=n_\mu^{\prime}(z) \, \diff z$. It is normalised such that $\int \diff \chi n_\mu(\chi) = 1$.

\section{Quadratic estimator} 
\label{sec:qe}

For the direct extraction of the shear power spectrum from the data one can for example use a maximum-likelihood technique employing a quadratic estimator \citep{Bond1998, Seljak1998, Hu2001} or measure a pseudo power spectrum from the Fourier-transformed shear field (also pseudo-$C(\ell)$; \citealt{Hikage2011, Asgari2016}). 
The likelihood-based quadratic estimator automatically accounts for any irregularity in the survey geometry or data sampling while it still maintains an optimal weighting of the data. This is important when dealing with real data because it allows for the use of sparse sampling techniques and it can deal efficiently with (heavily) masked data \citep{Koehlinger2016, Asgari2016}. 
A particular disadvantage of the quadratic estimator is that it requires an accurate and precise estimate of the noise in the data for the clean extraction of E- and B-modes. This is a very important point especially for current surveys in which the noise power dominates over the cosmological signal even on the largest scales. 
The pseudo-$C(\ell)$ method is faster thanks to efficient fast Fourier transforms, but in order to obtain an unbiased measurement of the shear power spectrum it requires a non-trivial deconvolution of the extracted pseudo spectrum with a window matrix. This deconvolution may lead to less accurate measurements on large scales \citep{Asgari2016}. 

Alternative pseudo-$C(\ell)$ methods are based on correlation-function measurements as input (e.g. \citealt{Schneider2002, Becker2015}). These present a hybrid approach, translating the real-space measurements and all their properties into Fourier-space, while formally requiring knowledge of the correlation-function measurements over all angles from zero to infinity.
Moreover, correlation-function based power spectrum estimators/translators rely on a non-trivial correction of the additive shear bias which is not required for the quadratic estimator as will be shown in Appendix~\ref{sec:c-term}.

\subsection{Method}

Here we only briefly summarise the quadratic-estimator algorithm applied to cosmic shear including its extension to tomographic bins. For an in-depth description we refer the reader to the original literature \citep{Hu2001, Lin2012, Koehlinger2016}.

\subsubsection{Likelihood}
\label{sec:shear_lklhood}

The likelihood of the measured shear field is assumed to be Gaussian over all scales of interest for our analysis, i.e.
\begin{equation}
\mathcal{L} = \frac{1}{(2 \upi)^N|\mathbfss{C}(\bmath{\mathcal{B}})|^{1/2}}\exp{ \left[-\tfrac{1}{2} \, \bmath{d}^T[\mathbfss{C}(\bmath{\mathcal{B}})]^{-1} \bmath{d} \right]} \, .
\end{equation}
The data vector $\bmath{d}$ with components
\begin{equation}
d_{a i \mu} = \gamma_{a}(\bmath{n}_i, \ z_\mu) 
\end{equation}
contains both components of the measured shear $\gamma_{a}$ per pixel $\bmath{n}_i$ for each redshift bin $z_\mu$. 
The covariance matrix $\mathbfss{C}$ is written as the sum of the noise $\mathbfss{C}^{\rm noise}$ and the cosmological signal $\mathbfss{C}^{\rm sig}$ (equation~\ref{eq:shear_corr_matrix}). The latter depends on the shear power spectra $C(\ell)$ which are approximated in the algorithm as piece-wise constant band powers $\bmath{\mathcal{B}}$. 

As long as the pixel noise of the detector is uncorrelated, the noise matrix can be assumed to be diagonal, i.e. shape noise is neither correlated between different pixels $\bmath{n}_i$, $\bmath{n}_j$ and shear components $\gamma_a$, $\gamma_b$, nor between different redshift bins $z_\mu, \ z_\nu$: 
\begin{equation} \label{eq:noise}
\mathbfss{C}^{\mathrm{noise}} = \frac{\sigma_{\tilde{\gamma}}^2(z_\mu)}{N_i(z_\mu)} \delta_{ij} \delta_{ab} \delta_{\mu\nu} \, , 
\end{equation}
where $\sigma_{\tilde{\gamma}}$ is the standard deviation of an unbiased shear estimator. Usually it is assumed that $\sigma_{\tilde{\gamma}} = \sigma_{\epsilon}$, the root-mean-square ellipticity per ellipticity component for all galaxies in the survey. $N_i(z_\mu)$ denotes the effective number of galaxies per pixel $i$ in redshift bin $z_\mu$.\footnote{The effective number of galaxies per pixel can be calculated using equation~(\ref{eq:n_eff}) multiplied by the area of the pixel $\Omega$.} The specification of the noise matrix here is one of the fundamental differences with respect to correlation-function measurements: whereas this algorithm requires a characterisation of the noise in the data \textit{before} performing the measurement, correlation-function measurements can be performed regardless of any knowledge of the noise. The decomposition of signal and noise enters then only in the covariance matrix of the real-space measurements.

As for current surveys the signal is still much weaker than the noise even at the lowest multipoles, an accurate and precise estimate of the noise level is paramount for an unbiased interpretation of the cosmological signal. 

This is difficult to achieve because the measured ellipticity dispersion, calculated as a weighted variance of galaxy ellipticities is a biased estimate of the shear dispersion.
We can understand this as arising from noise bias: for example galaxies with low signal-to-noise ratio (SNR) have broad likelihood surfaces which are biased to low ellipticity values and hence also to low ellipticity dispersion. The multiplicative bias correction (see Section~\ref{sec:m-correction} for a definition and \citealt{Fenech-Conti2016}) is derived for \textit{shear} from an ensemble of galaxies rather than \textit{ellipticity} measurements for individual galaxies. This allows us to derive an unbiased ensemble shear based on ellipticity measurements (see Section~\ref{sec:m-correction}), but it is not expected to correctly predict the bias on the ellipticity dispersion. Deriving a calibration for the shear dispersion is beyond the scope of this paper, but the impact of that will be scrutinised in Section~\ref{sec:cosmo_inference}.

In principle, the uncertainty in the noise level can be overcome by marginalising over one or more free noise amplitudes for each tomographic bin while extracting the data. However, \citet{Lin2012} observed that the simultaneous extraction of B-modes and a free noise amplitude is very challenging for noisy data. We therefore follow \citet{Lin2012} by fixing the noise properties to the measured values (Table~\ref{tab:n_eff}) while extracting E- and B-modes simultaneously.

\subsubsection{Maximum likelihood solution}
\label{sec:maxlkl_solution}

The best-fitting band powers $\mathcal{B}$ and the cosmic signal matrix $\mathbfss{C}^{{\rm sig}}$ that describe the measured shear data the best are found by employing a Newton--Raphson optimization. This algorithm finds the root of $\diff \mathcal{L}/ \diff \mathcal{B} = 0$ \citep{Bond1998, Seljak1998}, i.e. its maximum-likelihood solution, by iteratively stepping through the expression $\mathcal{B}_{\rm i+1} = \mathcal{B}_{\rm i} + \delta \mathcal{B}$ until it converges to the maximum-likelihood solution. 

With appropriate choices for an initial guess of the band powers and the step size parameter of the Newton--Raphson optimization, the method usually converges quickly towards the maximum-likelihood solution.
\citet{Hu2001} gave several empirical recommendations for a numerically stable and quick convergence. The most important one is to reset negative band powers to a small positive number at the start of an iteration. As a result a small bias is introduced in the recovered power spectrum, which depends on the amplitude of the signal (the closer the signal is to zero the larger is the overall effect) and on the noise level (the larger the noise the more often the resetting will occur). This `resetting bias' can be easily calibrated using mock data as shown in Section~\ref{sec:fiducial_B_modes}.

\subsubsection{Band window matrix}
\label{sec:bwm}

Each measured band power $\mathcal{B}$ samples the corresponding power spectrum with its own window function. For a general estimator we can relate the expectation value of the measured band power $\langle \mathcal{B} \rangle$ to the shear power spectrum $C$ at integer multipoles through the band-power window function $W(\ell)$ (\citealt{Knox1999, Lin2012}), i.e.   
\begin{equation} \label{eq:conv_window_func}
\langle \mathcal{B}_{\zeta\vartheta\beta} \rangle = \sum_\ell \tfrac{\ell(\ell + 1)}{2\pi} W_{(\zeta\vartheta\beta)(\zeta\vartheta)}(\ell)  C_{\zeta\vartheta}(\ell) \, , 
\end{equation}
where $W_{(\zeta\vartheta\beta)(\zeta\vartheta)}(\ell)$ denotes the elements of the block diagonal of the band window matrix $\mathbfss{W}(\ell)$. The index $\zeta$ labels the unique $n_z (n_z + 1) / 2$ redshift-bin correlations, the index $\vartheta$ the band power type (i.e. EE, BB, or EB), and the index $\beta$ runs over the band power bin, i.e. over a given range of multipoles. Equation~\ref{eq:conv_window_func} is required for inferring cosmological parameters from the measured band powers (see Section~\ref{sec:cosmo_inference}), because it translates a smooth cosmological signal prediction into band powers. Moreover, the full band window matrix $\mathbfss{W}(\ell)$ is required for propagating the properties of the quadratic estimator into the analytical covariance (see Section~\ref{sec:cov}). Note that due to the latter the notation in equation~(\ref{eq:conv_window_func}) has changed with respect to the one presented in \citet{Koehlinger2016}. We present the updated notation in Appendix~\ref{app:BWM_update}.

The sum is calculated for integer multipoles $\ell$ in the range $10 \leq \ell \leq 3000$ since the cosmological analysis uses multipoles in the range $76 \leq \ell \leq 1310$ (see Section~\ref{sec:data_meas}). Therefore, the lowest multipole for the summation should extend slightly below $\ell_{\rm field} = 76$ and the highest multipole should include multipoles beyond $\ell = 1310$ in order to capture the full behaviour of the band window function below and above the lowest and highest bands, respectively. 


Our technical implementation of the quadratic estimator algorithm employs the {\scriptsize NUMPY} package for {\scriptsize PYTHON}. This allows for performing calculations with 64-bit floating point precision. The inversion of the full covariance matrix, i.e. the sum of equations~(\ref{eq:shear_corr_matrix})~and~(\ref{eq:noise}) is performed once per Newton--Raphson iteration (although occurring multiple times in there, see e.g. equation~11 in \citealt{Koehlinger2016}). For the inversion we use the standard inversion routine from the linear algebra sub-package of {\scriptsize NUMPY}.\footnote{Version number 1.9.0., compiled with the Intel$^{\copyright}$ Math Kernel Library ({\scriptsize MKL}), version number 11.0.4.}. This routine in turn uses a linear equation solver employing an LU decomposition algorithm to solve for the inverse of the matrix. The inverse matrices of the largest matrices used in the subsequent analysis (i.e. ${\rm dim}(\bmath{\rm C}) \leq 9352^2$ for 2 z-bins and ${\rm dim}(\bmath{\rm C}) \leq 13998^2$ for 3 z-bins) pass the accuracy test of \citet{Newman1974}. Moreover, we verify that $| \bmath{\rm Id} - \bmath{\rm C} \bmath{\rm C}^{-1} |_{ij} \leq 10^{-14}$ for all elements $i$, $j$ of the matrices.

\subsection{Testing and calibration}
\label{sec:fiducial_B_modes}

For convergence and performance reasons, negative band powers are reset to a small positive number at the start of each iteration towards the maximum-likelihood solution. This procedure does not prevent the algorithm to yield negative band powers at the end of a Newton--Raphson iteration (as might be necessary due to noise), but it introduces a bias in the extracted band powers. The amplitude of the bias depends on the width of the band-power distribution which is set by the noise level in the data. Hence, a distribution of band powers expected to be centred around zero such as B-modes will be more biased than a distribution centred around a non-zero mean such as E-modes. The dependence of the bias on the noise level in the data can be characterised by using mock data in which the E- and B-modes are perfectly known. We use here a suite of B-mode free Gaussian Random Fields (GRFs) described in more detail in \citet{Koehlinger2016}.

\begin{figure*}
	\centering
	\includegraphics[width=180mm]{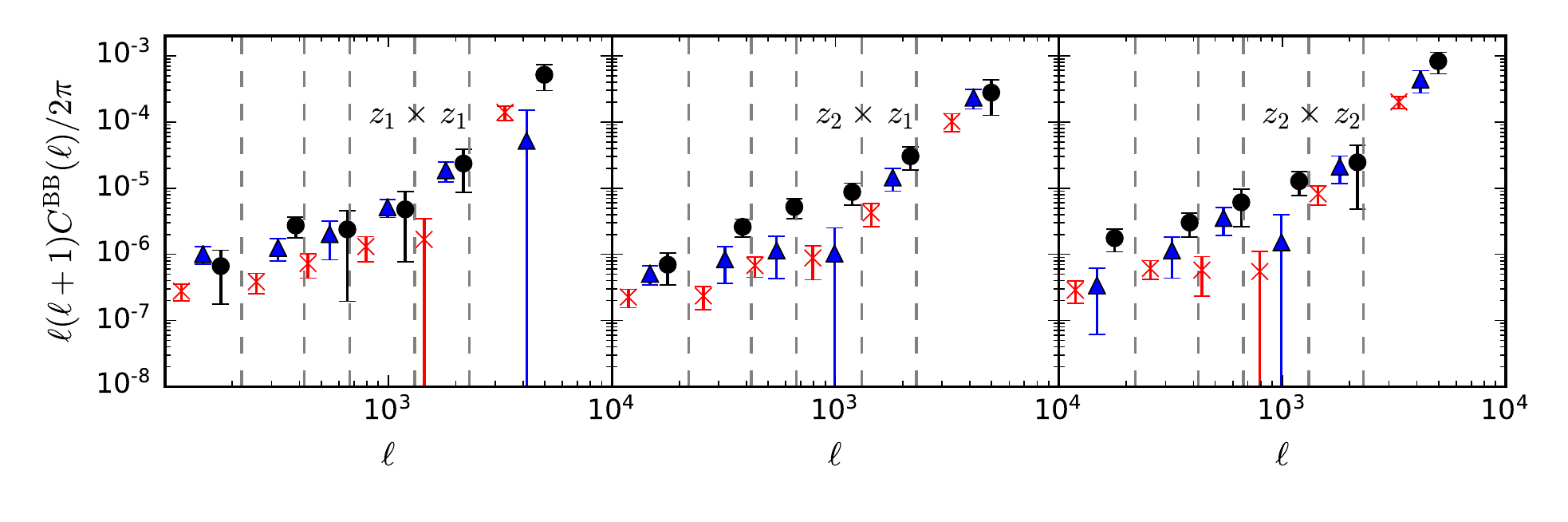}
	\caption{Extracted B-mode band powers as a function of multipole and redshift correlation (from left to right) from 50 Gaussian random field realizations for three different noise levels each. Crosses ({\color{red}red}) correspond to $\sigma_\epsilon = 0.10$, triangles ({\color{blue}blue}) to $\sigma_\epsilon = 0.19$, and circles (black) to $\sigma_\epsilon = 0.28$ for fixed number densities of $n_{\rm eff}(z_1) = 2.80 \, {\rm arcmin}^{-2}$ and $n_{\rm eff}(z_2) = 2.00 \, {\rm arcmin}^{-2}$. Crosses and circles are plotted with constant multiplicative offset in multipoles for illustrative purposes. The vertical dashed lines (grey) indicate the borders of the band power intervals (Table~\ref{tab:bp_intervals}). The errors are derived from the run-to-run scatter and divided by $\sqrt{50}$ to represent the error on the mean.}
	\label{fig:signals_BB_2zbins_GRFs}
\end{figure*}

We extract E- and B-modes simultaneously for three sets of 50 GRF realisations with varying noise levels (i.e. $\sigma_\epsilon = 0.10$, $\sigma_\epsilon = 0.19$, and $\sigma_\epsilon = 0.28$ for fixed $n_{\rm eff}(z_1) = 2.80 \, {\rm arcmin}^{-2}$, and $n_{\rm eff}(z_2) = 2.00 \, {\rm arcmin}^{-2}$). Each GRF field uses the survey mask of the CFHTLenS W2 field ($\approx 22.6 \, {\rm deg}^2$), which is an adequate representation of the KiDS subpatches (Section~\ref{sec:data_meas}) in terms of size and shape. For the extraction of the band powers we use the same multipole binning and shear pixel size employed in the subsequent KiDS-450 data extraction (see Section~\ref{sec:data_meas}). 
Although the GRFs are B-mode free by construction, Fig.~\ref{fig:signals_BB_2zbins_GRFs} shows significant extracted B-modes as expected. Moreover, the fact that the sets of extracted B-modes scale with the noise level built into the GRFs indicates that they are indeed caused by the noise-dependent `resetting bias'. In Figs.~\ref{fig:BWM_EE_z1z1_conv},~\ref{fig:BWM_EE_z2z1_conv},~and~\ref{fig:BWM_EE_z2z2_conv} from Appendix~\ref{app:power_leakage} we show explicitly that any contribution to these B-modes due to power leakage/mixing introduced by e.g. the survey mask are negligible. 

\begin{figure*}
	\centering
	\includegraphics[width=180mm]{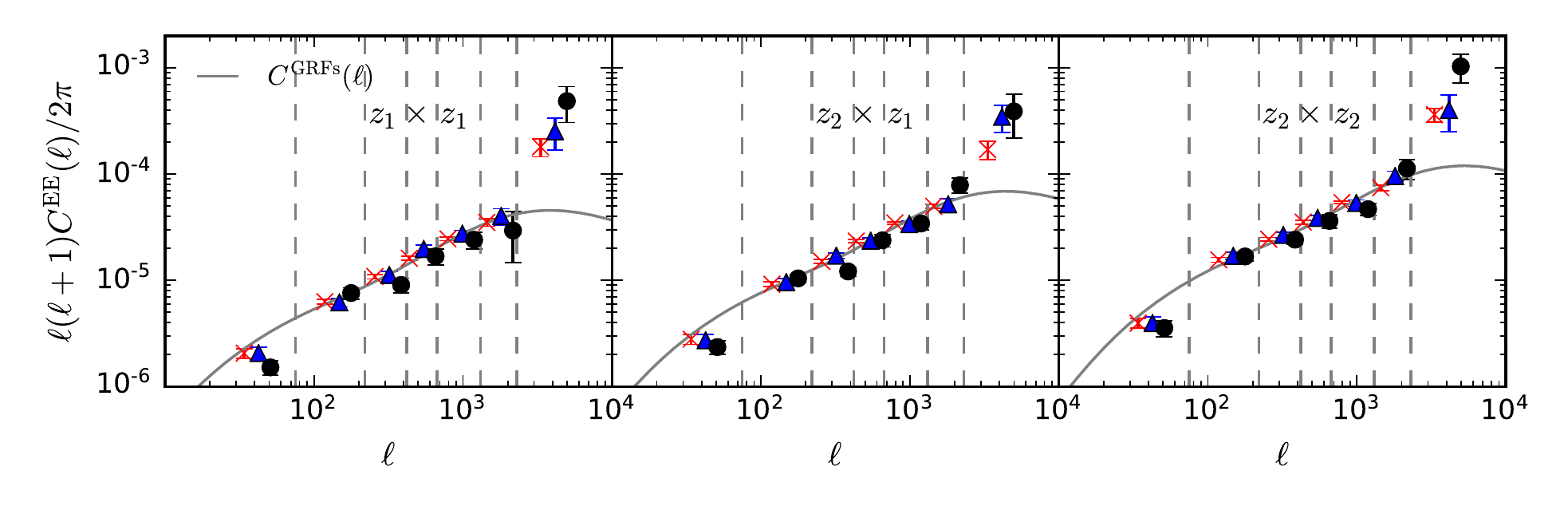}
	\caption{The same as in Fig.~\ref{fig:signals_EE_2zbins_GRFs} but for E-modes. The grey solid line in each panel shows the input power spectrum used for the creation of the Gaussian random fields (GRFs). A quantitative comparison between input power and extracted power for the highest noise sample is presented in Fig.~\ref{fig:comparison_E_modes_GRFs}. Note that the first and last band powers are not expected to recover the input power (Section~\ref{sec:band_power_select}).}	
	\label{fig:signals_EE_2zbins_GRFs}
\end{figure*}

The `resetting bias' will affect band powers whose distribution is expected to be centred around zero more strongly than band powers with a positive non-zero mean, therefore the impact of the bias on the extracted E-modes is expected to be negligible. This is indeed the case as the extracted E-modes in Fig.~\ref{fig:signals_EE_2zbins_GRFs} do not show a significant dependence on the noise level built into the GRFs except for the last band. For the second-to-last band in the highest noise realisation, however, there appears to be a bias, too. As we show in Fig.~\ref{fig:comparison_E_modes_GRFs} from Appendix~\ref{app:extra_fig} the input-power of the second-to-last band is still recovered within its $2\sigma$ error on the mean (whereas bands 1 to 5 are recovered within their $1\sigma$ errors on the mean). 
  
The explanation for this bias can be found in Fig.~\ref{fig:signals_EE_2zbins_GRFs}: if we focus on the second-to-last band, we notice that in the low-noise cases the extracted values are unbiased, while a deviation from the expected value is visible for the high-noise case (which is set to match the noise level of the data). For the other bands the extracted power is independent from the noise level. 
This noise dependence of the bias points to a degradation in the convergence of the Newton--Raphson method (for a fixed number of iterations) when the SNR of the data is very low, as noticed already by \citet{Hu2001}.

We further note that the errors on the mean derived from the 50 GRF runs on fields each of the size of W2 correspond effectively to those of a survey of about three times the size of the effective area used in KiDS-450. Therefore, the bias in the second-to-last band is expected to be negligible for the real data extraction. Nevertheless, we make the conservative choice of excluding the second-to-last band in the subsequent cosmological analysis (see Section~\ref{sec:band_power_select}).

With the three sets of simultaneously extracted E- and B-modes for varying noise properties of the GRFs, we derive a model for the fiducial B-modes caused by the `resetting bias'. All sets of band powers are modelled as a function of the noise with a power-law of the form:
\begin{align}
\label{eq:power_law_B_modes}
p_{{\rm rb}}(x) &= A_{{\rm rb}} x^{\beta_{{\rm rb}}} \, {\rm with} \\ 
x &= \frac{\ell (\ell + 1)}{2\pi } \frac{\sigma_\epsilon (z_\mu) \sigma_\epsilon (z_\nu)}{\sqrt{n(z_\mu)n(z_\nu)}} \, . \nonumber
\end{align}
Here, the variable $x$ encodes the implicit multipole and redshift dependencies. Note though that the multipole dependence is just an artefact of extracting the band powers with the normalisation $\ell (\ell +1) / 2 \pi$. 
We determine $A_{{\rm rb}} = (9.08 \pm 4.23) \times 10^{-4}$ and $\beta_{{\rm rb}} = 0.64 \pm 0.04$ by simultaneously fitting the power-law model to the sets of B-mode band powers.
The power-law model is also included in the cosmological likelihood code for a simultaneous evaluation of the E-mode and B-mode band powers to allow for a consistent error propagation through marginalising over the parameters $A_{\rm rb}$ and $\beta_{\rm rb}$. The details of this are given in Section~\ref{sec:other_sys}.

\section{Data: K\lowercase{i}DS-450}
\label{sec:data_meas}

In the following analysis we use the KiDS-450 dataset.
KiDS is an ongoing ESO optical survey which will eventually cover $1350 \deg^2$ in four bands ($u$, $g$, $r$, and $i$). It is carried out using the OmegaCAM CCD mosaic camera mounted at the Cassegrain focus of the VLT Survey Telescope (VST). The combination of camera and telescope was specifically designed for weak lensing studies and hence results in small camera shear and an almost round and well-behaved point spread function (PSF). The data processing pipeline from individual exposures in multiple colours to photometry employs the A{\scriptsize STRO}-WISE system \citep{Valentijn2007, Begeman2013}. For the lensing-specific data reduction of the $r$-band images, we use {\scriptsize THELI} \citep{Erben2005, Erben2009, Erben2013, Schirmer2013}. The galaxy shapes are measured from the {\scriptsize THELI}-processed data with the shape measurement software \lensfit \ \citep{Miller2013, Fenech-Conti2016}. The full description of the pipeline for previous data releases of KiDS (DR1/2) is documented in \citet{deJong2015} and \citet{Kuijken2015}. All subsequent improvements applied to the data processing for KiDS-450 are summarised in \citet{Hildebrandt2016}. The \lensfit -specific updates including a description of the extensive image simulations for shear calibrations at the sub-percent level are documented in \citet{Fenech-Conti2016}.

The interpretation of the cosmic shear signal also requires accurate and precise redshift distributions, $n(z)$ (equation~\ref{eq:lensing_kernel}). For the estimation of individual photometric redshifts for source galaxies the code {\scriptsize BPZ} \citep{Benitez2000} is used following the description in \citet{Hildebrandt2012}. In earlier KiDS and CFHTLenS analyses the overall $n(z)$ was used based on the stacked redshift probability distributions of individual galaxies, $p(z)$, as estimated by {\scriptsize BPZ}. However, as shown in \citet{Hildebrandt2016, Choi2015} the $n(z)$ estimate in this way is biased at a level that is intolerable for current and especially future cosmic shear studies (see \citealt{Newman2015, Choi2015} for a discussion).  

\citet{Hildebrandt2016} employed a weighted direct calibration (`DIR') of photometric redshifts with spectroscopic redshifts. This calibration method uses several spectroscopic redshift catalogues from surveys overlapping with KiDS.
In practice, spectroscopic redshift catalogues are neither complete nor a representative sub-sample of the photometric redshift catalogues currently used in cosmic shear studies. In order to alleviate these practical shortcomings the photometric redshift distributions and the spectroscopic redshift distributions are re-weighted in a multi-dimensional magnitude-space, so that the volume density of objects in this magnitude space matches between photometric and spectroscopic catalogues \citep{Lima2008}. The direct calibration is further cross-checked with two additional methods and found to yield robust and accurate estimates of the photometric redshift distribution of the galaxy source sample (see \citealt{Hildebrandt2016} for details).

The fiducial KiDS-450 dataset consists of 454 individual ${\sim} 1 \deg^2$ tiles (see fig.~1 from \citealt{Hildebrandt2016}). The $r$-band is used for the shape measurements with a median and maximum seeing of $0.66 \, {\rm arcsec}$ and $0.96 \, {\rm arcsec}$, respectively. The tiles are grouped into five patches (and corresponding catalogues) covering an area of ${\approx} 450 \deg^2$ in total. 
After masking stellar haloes and other artefacts in the images, the total area of KiDS-450 is reduced to an effective area usable for lensing of about $360 \deg^2$. Since the catalogue for an individual KiDS patch contains long stripes (e.g. $1 \deg$ by several degrees) or individual tiles due to the pointing strategy, we exclude these disconnected tiles from our analysis, which amounts to a reduction in effective area by ${\approx} 36 \deg^2$ compared to \citet{Hildebrandt2016}. Moreover, the individual patches are quite large resulting in long runtimes for the signal extraction. Therefore, we split each individual KiDS patch further into two or three subpatches yielding 13 subpatches in total with an effective area of $323.9 \deg^2$. Each subpatch contains a comparable number of individual tiles. The splitting into subpatches was performed along borders that do not split individual tiles, as a single tile represents the smallest data unit for systematic checks and further quality control tests. 

The coordinates in the catalogues are given in a spherical coordinate system measured in right ascension $\alpha$ and declination $\delta$. Before we pixelize each subpatch into shear pixels, we first deproject the spherical coordinates into flat coordinates using a tangential plane projection (also known as gnomonic projection). The central point for the projection of each subpatch, i.e. its tangent point, is calculated as the intersection point of the two great circles spanned by the coordinates of the edges of the subpatch. 

The shear components $g_a$ per pixel at position $\bmath{n}=(x_c, y_c)$ are estimated from the ellipticity components $e_a$ inside that pixel:
\begin{equation}
g_a(x_{\rm c}, y_{\rm c}) = \frac{\sum_{i} w_i e_{a, i}}{\sum_{i} w_i} \, ,
\end{equation}
where the index $a$ labels the two shear and ellipticity components, respectively, and the index $i$ runs over all objects inside the pixel. The ellipticity components $e_a$ and the corresponding weights $w$ are computed during the shape measurement with \lensfit \ and they account both for the intrinsic shape noise and measurement errors. 

For the position of the average shear we take the centre of the pixel (hence the subscript `c' in the coordinates). Considering the general width of our multipole band powers it is justified to assume that the galaxies are uniformly distributed in each shear pixel. Finally, we define distances $r_{ij}=|\bmath{n}_i-\bmath{n}_j|$ and angles $\varphi=\arctan{(\Delta y/\Delta x)}$ between shear pixels $i$, $j$ which enter in the quadratic estimator algorithm (see Section~\ref{sec:qe}). 

\subsection{Band power selection}
\label{sec:band_power_select}

The lowest scale of the multipole band powers that we extract is in general set by the largest separation $\theta_{\rm max}$ possible between two shear pixels in each subpatch. In a square-field that would correspond to the diagonal separation of the pixels in the corners of the patch. However, this would yield only two independent realisations of the corresponding multipole $\ell_{\rm min}$. Instead, we define the lowest physical multipole $\ell_{\rm field}$ as corresponding to the distance between two pixels on opposite sides of the patch ensuring that there exist many independent realisations of that multipole so that a measurement is statistically meaningful. 

In general, the subpatches used in this analysis are not square but rectangular and hence we follow the conservative approach of defining $\ell_{\rm field}$ corresponding to the shorter side length of the rectangle. 
The shortest side length is $\theta \approx 4\fdg 74$ corresponding to $\ell_{\rm field} = 76$. 

The lowest multipole over all subpatches is $\ell_{\rm min} = 34$ corresponding to a distance $\theta \approx 10\fdg 5$, but we set the lower border of the first band power even lower to $\ell = 10$. That is because the quadratic estimator approach allows us to account for any leftover DC offset\footnote{Signal processing terminology in which DC refers to direct current.}, i.e. a non-zero mean amplitude, in the signal by including even lower multipoles than $\ell_{\rm min}$ in the first band power (see Appendix~\ref{sec:c-term}).

The highest multipole $\ell_{\rm max}$ available for the data analysis is set by the side length of the shear pixels. The total number of shear pixels in the analysis is also a critical parameter for the runtime of the algorithm because it sets the dimensionality of the fundamental covariance matrix (equation~\ref{eq:shear_corr_matrix}), together with the number of redshift bins and the duality of the shear components. Moreover, Gaussianity is one of the assumptions behind the quadratic estimator which naturally limits the highest multipole to the mildly non-linear regime \citep{Hu2001}. 
Hence, we set $\sigma_{\rm pix} = 0\fdg 12$ corresponding to a maximum multipole $\ell_{\rm pix} = 3000$. At the median redshift of the survey, $z_{\rm med} = 0.62$, this corresponds to a wavenumber $k = 1.89 \, h {\rm Mpc}^{-1}$.

The borders of the last band should however extend to at least $2 \ell_{\rm pix} \approx 6000$ due to the increasingly oscillatory behaviour of the pixel window function (equation~\ref{eq:pixel_window}) close to and beyond $\ell_{\rm pix}$. The width of all intermediate bands should be at least $2 \ell_{\rm field}$ in order to minimize the correlations between them \citep{Hu2001}. 
Given all these constraints we extract in total seven E-mode band powers over the range $10 \leq \ell \leq 6000$. 

For the cosmological analysis we will drop the first, second-to-last and last band powers. The first band power is designed to account for any remaining DC offset in the data (see Appendix~\ref{sec:c-term}) and should therefore be dropped. The last band power sums up the oscillating part of the pixel window and should also be dropped. As noted already in Section~\ref{sec:fiducial_B_modes}, tests on GRF mock data showed that the input power for the second-to-last band is only recovered within its $2\sigma$ error bar (see Fig.~\ref{fig:comparison_E_modes_GRFs}). Therefore, we make the conservative choice of excluding the second-to-last band in addition to the first and last band in the subsequent cosmological analysis (also taking into account its low SNR). We confirmed though that including the second-to-last E-mode band power (and its corresponding B-mode) does not change the conclusions of the cosmological inference (Section~\ref{sec:cosmo_inference}).

In addition to the E-modes, we simultaneously extract six B-mode band powers. Their multipole ranges coincide with the ranges of the E-mode bands 2 to 7. The lowest multipole band is omitted because on scales comparable to the field size, the shear modes can no longer be split unambiguously into E- and B-modes.
All ranges are summarised in Table~\ref{tab:bp_intervals} where we also indicate the corresponding angular scales. Note, however, that the na\"ive conversion from multipole to angular scales is insufficient for a proper comparison to correlation function results. An outline of how to compare both approaches properly is given in Appendix~\ref{sec:theta_comparison}.     

\begin{table}
	\caption{Band-power intervals.}
	\label{tab:bp_intervals}
	\begin{center}
		\begin{tabular}{ c c c c }
			\toprule 
			Band No.& $\ell$--range& $\theta$--range& Comments\\
			\midrule 
			1& 10--75& $2160.0$--$288.0 \, {\rm arcmin}$& (a), (b)\\
			2& 76--220& $284.2$--$98.2 \, {\rm arcmin}$& --\\
			3& 221--420& $98.0$--$51.4 \, {\rm arcmin}$& --\\
			4& 421--670& $51.3$--$32.2 \, {\rm arcmin}$& --\\
			5& 671--1310& $32.2$--$16.5 \, {\rm arcmin}$& --\\
			6& 1311--2300& $16.5$--$9.4 \, {\rm arcmin}$& (a)\\
			7& 2301--6000& $9.4$--$3.6 \, {\rm arcmin}$& (a)\\
			\bottomrule 
		\end{tabular}
	\end{center}
	\medskip 
	\textit{Notes.} (a) Not used in the cosmological analysis. (b) No B-mode extracted. The $\theta$-ranges are just an indication and cannot be compared directly to $\theta$-ranges used in real-space correlation function analyses due to the non-trivial functional dependence of these analyses on Bessel functions (see Appendix~\ref{sec:theta_comparison}).
\end{table}

We calculate the effective number density of galaxies used in the lensing analysis following \citet{Heymans2012} as
\begin{equation}
\label{eq:n_eff}
n_\mathrm{eff} = \frac{1}{\Omega}\frac{(\sum_i w_i)^2}{\sum_i w_i^2} \, ,
\end{equation} 
where $w$ is the \lensfit \ weight and the unmasked area is denoted as $\Omega$. In Table~\ref{tab:n_eff} we list the effective number densities per KiDS patch and redshift bin. Note that alternative definitions for $n_{\rm eff}$ exist, but this one has the practical advantage that it can be used directly to set the source number density in the creation of mock data. Moreover, equation~(\ref{eq:n_eff}) is the correct definition to use for analytic noise estimates.

\begin{table}
	\caption{Properties of the galaxy source samples.}
	\label{tab:n_eff}
	\begin{center}
	\resizebox{\columnwidth}{!}{%
		\begin{tabular}{ c c c c c c c c c }
			\toprule 
			redshift bin& $z_{\rm med}$& $N$& $n_{\rm eff}$& $\sigma_{\epsilon}$& $m_{\rm fid}(z_\mu)$\\
			\midrule 
			\textbf{2 z-bins:}\\
			$z_1$: $0.10 < z_{\rm B} \leq 0.45$& 0.41& $5\, 923\, 897$& $3.63$& 0.2895& $-0.013 \pm 0.010$\\
			$z_2$: $0.45 < z_{\rm B} \leq 0.90$& 0.70& $6\, 603\, 721$& $3.89$& 0.2848& $-0.012 \pm 0.010$\\
			\textbf{3 z-bins:}\\
			$z_1$: $0.10 < z_{\rm B} \leq 0.30$& 0.39& $3\, 879\, 823$& $2.35$& 0.2930& $-0.014 \pm 0.010$\\
			$z_2$: $0.30 < z_{\rm B} \leq 0.60$& 0.46& $4\, 190\, 501$& $2.61$& 0.2856& $-0.010 \pm 0.010$\\
			$z_3$: $0.60 < z_{\rm B} \leq 0.90$& 0.76& $4\, 457\, 294$& $2.56$& 0.2831& $-0.017 \pm 0.010$\\
			\bottomrule 
		\end{tabular}}
	\end{center}
	\medskip 
	\textit{Notes.} The median redshift $z_{\rm med}$, the total number of objects $N$, the effective number density of galaxies $n_\mathrm{eff}$ per arcmin$^2$ (equation~\ref{eq:n_eff}), the dispersion of the intrinsic ellipticity distribution $\sigma_\epsilon$, and fiducial multiplicative shear calibration $m_{\rm fid}$ per redshift bin for the KiDS-450 dataset used in our analysis.
\end{table}

\begin{figure*}
	\centering
	\includegraphics[width=180mm]{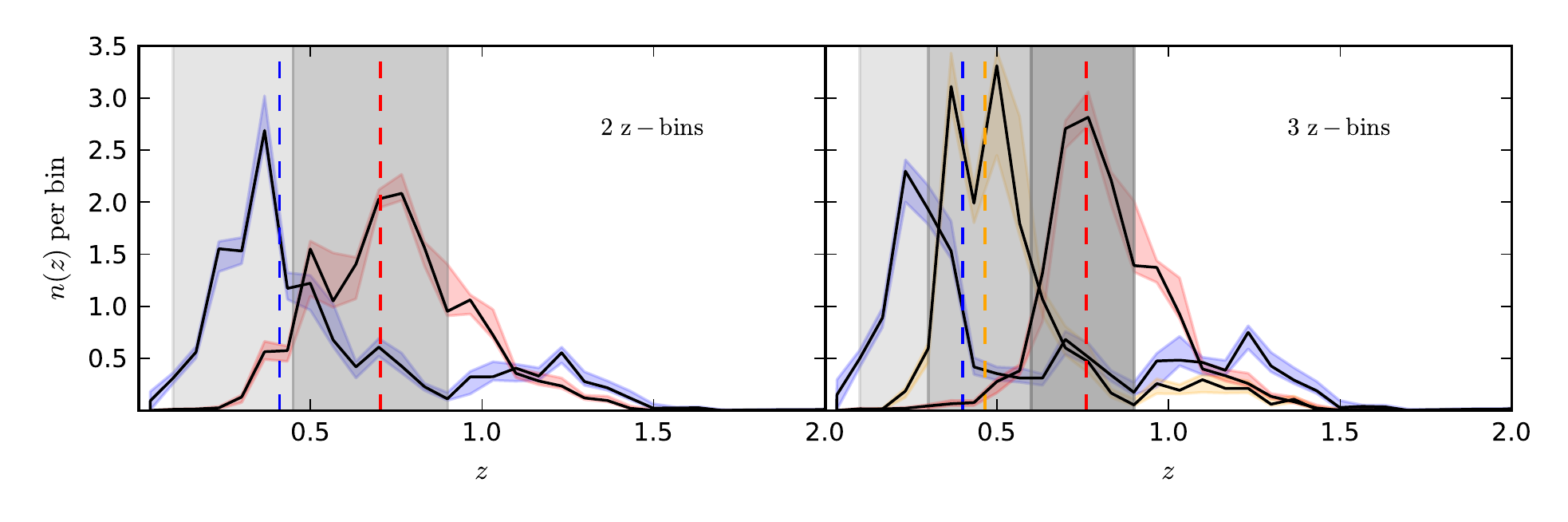}
    \caption{The normalised redshift distributions for the full sample, two and three tomographic bins employed in this study and estimated from the weighted direct calibration scheme (`DIR') presented in \citet{Hildebrandt2016}. The dashed vertical lines mark the median redshift per bin (Table~\ref{tab:n_eff}) and the (grey) shaded regions indicate the target redshift selection by cutting on the Bayesian point estimate for photometric redshifts $z_{\rm B}$. The (coloured) regions around each fiducial $n(z)$ per bin shows the $1\sigma$-interval estimated from $1000$ bootstrap realisations of the redshift catalogue. Lower panel: the summed and re-normalised redshift distribution over all tomographic bins.}
    \label{fig:n_of_z}
\end{figure*}

As discussed in \citet{Hildebrandt2016} the `DIR' calibration as well as the multiplicative shear bias corrections (Section~\ref{sec:m-correction}) are only valid in the range $0.10 < z_{\rm B} \leq 0.90$, where $z_{\rm B}$ is the Bayesian point estimate of the photometric redshifts from {\scriptsize BPZ} \citep{Benitez2000}. 
For the subsequent analysis we divide this range further into two and three tomographic bins with similar effective number densities (Table~\ref{tab:n_eff} and Fig.~\ref{fig:n_of_z}). Note that $z_{\rm B}$ is only used as a convenient quantity to define tomographic bins, but does not enter anywhere else in the analysis. 
The limitation to at most three redshift bins is due to runtime, since the dimension of the fundamental covariance matrix (equation~\ref{eq:shear_corr_matrix}) depends quadratically on the number of redshift bins, as noted earlier in this section. Applying the method also to only two redshift bins here serves as a cross-check of the 3 z-bin analysis. 

In Fig.~\ref{fig:n_of_z} we show the normalised redshift distributions for two and three redshift bins. The coloured regions around each $n(z)$ show the $1\sigma$-error estimated from $1000$ bootstrap realisations of the redshift catalogues per tomographic bin. This does not account for cosmic variance, but the effect on the derived $n(z)$ is expected to be small (see \citealt{Hildebrandt2016} for a discussion).

\subsection{Multiplicative bias correction}
\label{sec:m-correction}
The observed shear $\gamma_{\rm obs}$, measured as a weighted average of galaxy ellipticities, is generally a biased estimator of the true shear $\gamma$. The bias is commonly parametrized as \citep{Heymans2006}
\begin{equation}
\label{eq:observed_shear}
\gamma_{\rm obs} = (1+m)\gamma + c \, ,
\end{equation}  
where $m$ and $c$ refer to the multiplicative bias and additive bias, respectively.

The multiplicative bias is mainly caused by the effect of pixel noise in the measurements of galaxy ellipticities \citep{Melchior2012, Refregier2012, Miller2013}, but it can also arise if the model used to describe the galaxy profile is incorrect, or if stars are misclassified as galaxies. The latter two effects are generally subdominant compared to the noise bias. 
We quantify the amplitude of the multiplicative bias in the KiDS data by means of a dedicated suite of image simulations \citep{Fenech-Conti2016}. We closely follow the procedure described in there and derive a multiplicative correction for each tomographic bin as listed in Table~\ref{tab:n_eff}. The error bars account for statistical uncertainties and systematic errors due to small differences between data and simulations. In our likelihood analysis we apply the multiplicative correction to the measured shear power spectrum and its covariance matrix. In order to also marginalise over the uncertainties of this $m$-correction we propagate them into the likelihood analysis. 
As the errors on the $m_{\rm fid}(z_\mu)$ are fully correlated \citep{Fenech-Conti2016, Hildebrandt2016} we only need to include one free nuisance parameter per analysis. We apply the m-correction and propagate its uncertainty $\sigma_m = 0.01$ by varying a dummy variable $m$ within a flat $2\sigma_m$ prior centred on the fiducial value $m_{\rm fid}(z_1)$ for the first redshift bin in each step $i$ of the likelihood estimation. The value for each applied m-correction $m(z_\mu)$ is then fixed through the relation $m_i(z_\mu) = m_{\rm fid}(z_\mu) + \Delta m_i$ with $\Delta m_i = m - m_{\rm fid}(z_\mu)$. Hence, in the modelling of the power spectra for inferring cosmological parameters (Section~\ref{sec:cosmo_inference}) we include a nuisance parameter $m$ (Table~\ref{tab:n_eff}).

\subsection{Covariance}
\label{sec:cov}

An important ingredient for an accurate and precise inference of cosmological parameters from the measured band powers is the covariance matrix. There are several approaches to estimate the covariance matrix: the brute-force approach of extracting it directly from a statistically significant number (to reduce numerical noise) of mock catalogues, an analytical calculation or the inverse of the Fisher matrix calculated during the band-power extraction. Of course, each method has its specific advantages and disadvantages. The brute-force approach requires significant amounts of additional runtime, both for the creation of the mocks and the signal extraction. This can become a severe issue especially if the signal extraction is also computationally demanding, as is the case for the (tomographic) quadratic estimator. Moreover, if the mocks are based on $N$-body simulations the particle resolution and box size of these set fundamental limits for the scales that are available for a covariance estimation and to the level of accuracy and precision that is possible to achieve. 

In contrast, the Fisher matrix is computationally the cheapest estimate of the covariance matrix since it comes at no additional computational costs. However, it is only an accurate representation of the true covariance in the Gaussian limit and hence the errors for the non-linear scales will be underestimated. Moreover, the largest scale for a Fisher matrix based covariance is limited to the size of the patch. Therefore, the errors for scales corresponding to the patch size will also be underestimated. A possible solution to the shortcomings of the previous two approaches is the calculation of an analytical covariance matrix. This approach is computationally much less demanding than the brute-force approach and does not suffer from the scale-dependent limitations of the previous two approaches. Moreover, the non-Gaussian contributions at small scales can also be properly calculated. 

Hence, we follow the fiducial approach of \citet{Hildebrandt2016} and adopt their method for computing the analytical covariance (except for the final integration to correlation functions). The model for the analytical covariance consists of the following three components:\\
(i) a disconnected part that includes the Gaussian contribution to shape-noise, sample variance, and a mixed noise-sample variance term,\\
(ii) a non-Gaussian contribution from in-survey modes originating from the connected matter-trispectrum, and\\
(iii) a contribution from the coupling of in-survey and super-survey modes, i.e. super-sample covariance (SSC).\\

We calculate the first Gaussian term from the formula presented in \citet{Joachimi2008} employing the effective survey area $A_{\rm eff}$ (to take into account the loss of area through masking), the effective number density $n_{\rm eff}$ per redshift bin (to account for the \lensfit \ weights), and the weighted intrinsic ellipticity dispersion $\sigma_\epsilon$ per redshift bin (Table~\ref{tab:n_eff}). The required calculation of the matter power spectrum makes use of a `WMAP9' cosmology\footnote{$\Omega_{\rm m} = 0.2905$, $\Omega_\Lambda = 0.7095$, $\Omega_{\rm b} = 0.0473$, $h = 0.6898$, $\sigma_8 = 0.826$, and $n_{\rm s} = 0.969$ \citep{Hinshaw2013}}, the transfer functions by \citet{EisensteinHu1998}, and the recalibrated non-linear corrections from \citet{Takahashi2012}. 
Convergence power spectra are then calculated using equation~(\ref{eq:theo_power_spec}). 

The non-Gaussian `in-survey' contribution of the second term is derived following \citet{TakadaHu2013}. The connected trispectrum required in this step is calculated in the halo model approach employing both the halo mass function and halo bias from \citet{Tinker2008, Tinker2010}. For that we further assume an NFW halo profile \citep{NFW1997} with the concentration-mass relation by \citet{Duffy2008} and use the analytical form of its Fourier transform as given in \citet{Scoccimarro2001}. 

\citet{TakadaHu2013} model the final super-sample covariance (SSC) term as a response of the matter power spectrum to a background density consisting of modes exceeding the survey footprint. Again we employ the halo model to calculate this response. We note that in this context the corresponding contributions are also sometimes referred to as halo sample variance, beat coupling, and a dilation term identified by \citet{Li2014}. The cause for the coupling of super-survey modes into the survey is the finite survey footprint. For the proper modelling of this effect we create a {\scriptsize HEALPIX} \citep{Gorski2005} map of our modified KiDS-450 footprint (with $N = 1024$ pixels). Then the parts of the formalism by \citet{TakadaHu2013} related to survey geometry are converted into spherical harmonics. 

Based on the above description, we calculate the analytical covariance matrix $\mathbfss{C}_{(\zeta\vartheta)(\zeta^\prime \vartheta^\prime)}(\ell, \ell^\prime)$ at integer multipoles $\ell, \, \ell^\prime$ over the range $10 \leq \ell, \, \ell^\prime \leq 3000$\footnote{This range matches the range over which we later perform the summation when we convolve the theoretical signal predictions with the band window functions.} where the index pairs $\zeta, \, \zeta^\prime$ and $\vartheta, \, \vartheta^\prime$ label the unique redshift correlations and band types (EE and BB), respectively. Note that the EE to BB and vice versa the BB to EE part of this matrix is zero, i.e. there is no power leakage for an ideal estimator. Finally, we create the analytical covariance matrix of the measured band powers by convolving $\mathbfss{C}_{(\zeta\vartheta)(\zeta^\prime \vartheta^\prime)}(\ell, \ell^\prime)$ with the full band window matrix:  
\begin{equation}
\label{eq: ana_cov}
\mathbfss{C}_{AB} = \widetilde{\mathbfss{W}}_{A\zeta \vartheta}(\ell) \, \mathbfss{C}_{(\zeta\vartheta)(\zeta^\prime \vartheta^\prime)}(\ell, \ell^\prime) \, (\widetilde{\mathbfss{W}}^{\rm T})_{B\zeta^\prime \vartheta^\prime}(\ell^\prime) \, ,
\end{equation}
where the super-indices $A$, $B$ run over the band powers, their types (i.e. EE and BB), and the unique redshift correlations. $\widetilde{\mathbfss{W}}$ is the band window matrix defined in equation~(\ref{eq:window}) multiplied with the normalization for band powers, i.e. $\ell(\ell+1)/(2 \upi)$.
Note that through this matrix multiplication with the band window matrix all properties of the quadratic estimator are propagated into the band power covariance. 

\citet{Hildebrandt2016} presented a cross-check of the analytical covariance comparing it to numerical and jackknife covariance estimates. They found the analytical covariance to be a reliable, noise-free, and fast approach for estimating a covariance that includes SSC. Therefore, we use the analytical covariance here as our default, too.

\section{Shear power spectra from K\lowercase{i}DS-450}
\label{sec:signal}

For each of our 13 subpatches of the KiDS-450 data we extract the weak lensing power spectra in band powers spanning the multipole range $10 \leq \ell \leq 6000$ (see Section~\ref{sec:data_meas} and Table~\ref{tab:bp_intervals}). The measurements are performed in two and three redshift bins in the ranges listed in Table~\ref{tab:n_eff}. This yields in total $n_z (n_z + 1)/2$ unique cross-correlation spectra, including $n_z$ auto-correlation spectra per subpatch depending on the total number of z-bins, $n_z$. In the subsequent analysis we combine all spectra by weighting each spectrum with the effective area of the subpatch. This weighting is optimal in the sense that the effective area is proportional to the number of galaxies per patch and this number sets the shape noise variance of the measurements. 

\begin{figure*}
	\centering
	\includegraphics[width=180mm]{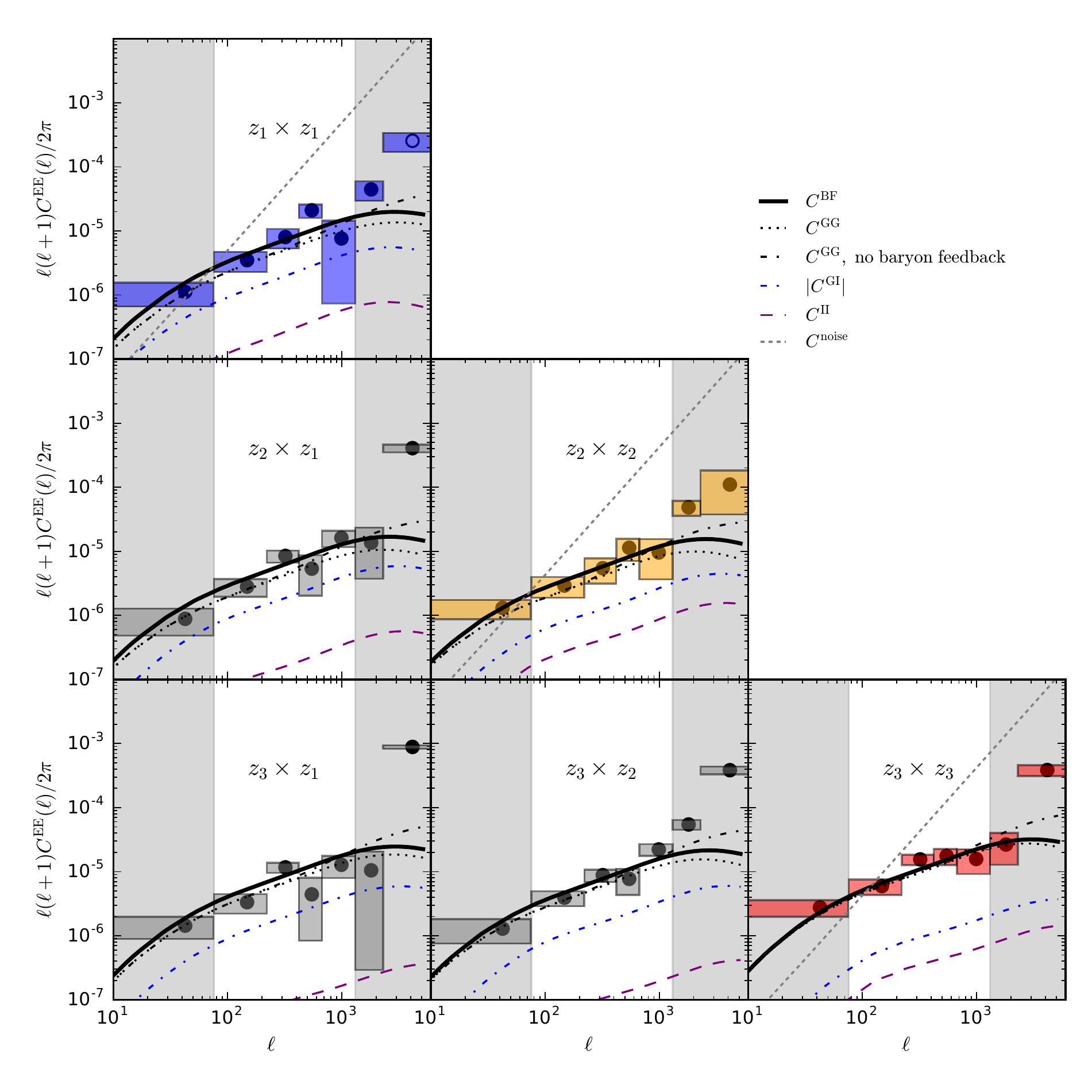}
	\caption{Measured E-mode band powers in three tomographic bins averaged with the effective area per patch over all 13 KiDS-450 subpatches. On the diagonal we show from the top-left to the bottom-right panel the auto-correlation signal of the low-redshift bin (blue), the intermediate-redshift bin (orange), and the high-redshift bin (red). The unique cross-correlations between these redshift bins are shown in the off-diagonal panels (grey). Note that negative band powers are shown at their absolute value with an open symbol. The redshift bins targeted objects in the range $0.10 < z_1 \leq 0.30$ for the lowest bin, $0.30 < z_2 \leq 0.60$ for the intermediate bin, and $0.60 < z_3 \leq 0.90$ for the highest bin. The $1\sigma$-errors in the signal are derived from the analytical covariance convolved with the averaged band window matrix (Section~\ref{sec:cov}), whereas the extent in $\ell$-direction is the width of the band. Band powers in the shaded regions (grey) to the left and right of each panel are excluded from the cosmological analysis (Section~\ref{sec:band_power_select}). The solid line (black) shows the power spectrum for the best-fitting cosmological model (Section~\ref{sec:cosmo_inference}). Moreover, we show the intrinsic alignment contributions, i.e. $C^{\rm GG}$ as dotted black line, $|C^{\rm GI}|$ as dash-dotted blue line, and $C^{\rm II}$ as dashed purple line. In addition to that, we also show $C^{\rm GG}$ without baryon feedback as a dashed black line. Note that for an accurate comparison of theory to data such as presented in Section~\ref{sec:cosmo_inference}), the theoretical power spectrum must be transformed to band powers (equation~\ref{eq:conv_window_func}). The dashed grey lines in the redshift auto-correlation models indicate the noise power spectrum in the data (Table~\ref{tab:n_eff}), which does not contribute to the redshift cross-correlations. Note, however, that the band powers are centred at the na\"ive $\ell$-bin centre and thus the convolution with the band window function is not taken into account in this figure, in contrast to the cosmological analysis.}
	\label{fig:signals_EE_3zbins}
\end{figure*}

We present the seven E-mode band powers for two and three redshift bins in Figs.~\ref{fig:signals_EE_3zbins}~and~\ref{fig:signals_EE_2zbins}.
The errors on the signal are estimated from the analytical covariance (Section~\ref{sec:cov}), which includes contributions from shape noise, cosmic variance, and super-sample variance. The width of the band is indicated by the  extent along the multipole axis. The signal is plotted at the na\"ive centre of the band, whereas for the subsequent likelihood analysis we take the window functions of the bands into account (equation~\ref{eq:conv_window_func}). 

In each redshift auto-correlation panel we show the average noise-power contribution calculated from the numbers in Table~\ref{tab:n_eff}. This noise component is removed from the data by the quadratic estimator algorithm yielding the band powers shown in Figs.~\ref{fig:signals_EE_3zbins}~(3~z-bins)~and~\ref{fig:signals_EE_2zbins}~(2~z-bins).
Only the bands outside the (grey) shaded areas enter in the cosmological analysis, i.e. we exclude the first, second-to-last and last band as discussed in Section~\ref{sec:band_power_select}. 

\begin{figure*}
	\centering
	\includegraphics[width=180mm]{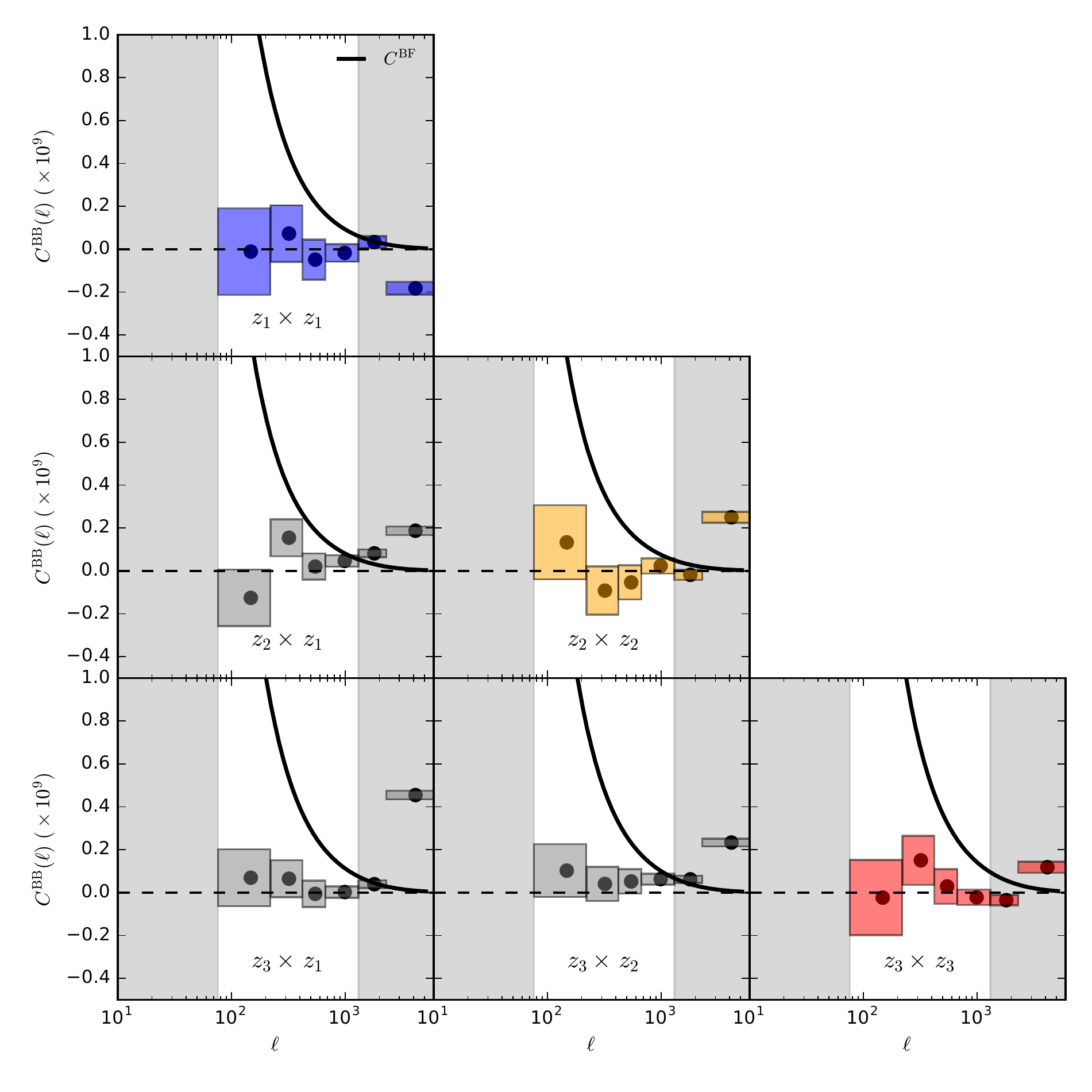}
	\caption{Same as Fig.~\ref{fig:signals_EE_3zbins} but for B-mode band powers corrected for the `resetting bias' introduced by the algorithm (Section~\ref{sec:fiducial_B_modes}). Note, however, the different scale (linear) and normalization used here with respect to Fig.~\ref{fig:signals_EE_3zbins}; for reference we also plot the best-fitting E-mode power spectrum as solid line (black). We show the measured B-modes as (black) dots with $1\sigma$-errors derived from the averaged shape-noise contribution to the analytical covariance convolved with the B-mode part of the averaged band window matrix. Note that the last band at high multipoles in each panel is designed to sum up the oscillating part of the pixel-window function and hence intrinsically biased.}
	\label{fig:signals_BB_3zbins}
\end{figure*}

\begin{figure*}
	\centering
	\includegraphics[width=180mm]{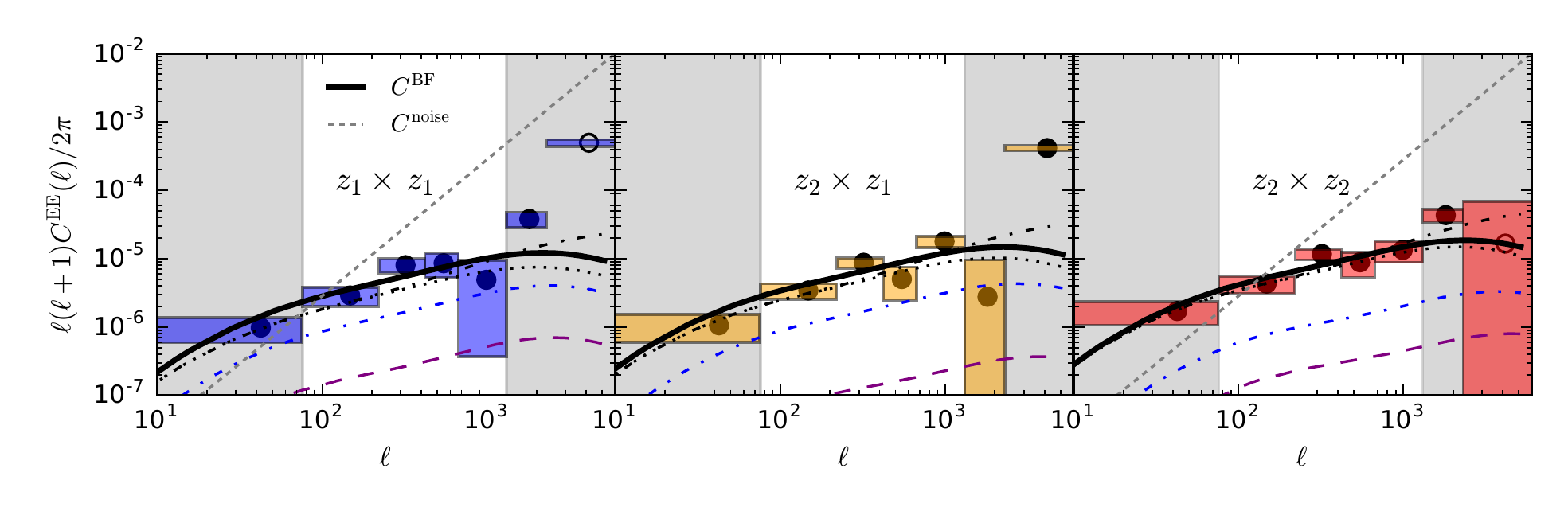}
	\caption{Same as Fig.~\ref{fig:signals_EE_3zbins} but for only two tomographic bins targeting objects with redshifts in the range $0.10 < z_1 \leq 0.45$ and $0.45<z_2\leq 0.90$. Please refer to the legend and caption of Fig.~\ref{fig:signals_EE_3zbins} for a full description of the theory components.} 
	\label{fig:signals_EE_2zbins}
\end{figure*}

\begin{figure*}
	\centering
	\includegraphics[width=180mm]{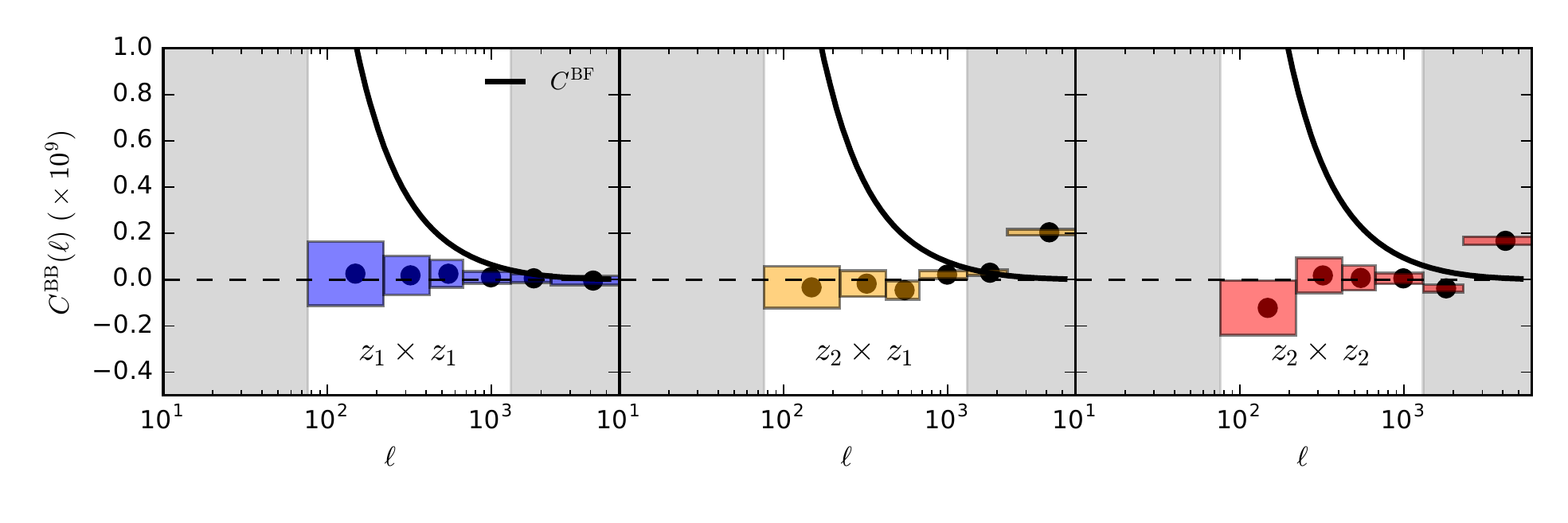}
	\caption{Same as Fig.~\ref{fig:signals_BB_3zbins} but for the same two tomographic bins used in Fig.~\ref{fig:signals_EE_2zbins}.}  	
	\label{fig:signals_BB_2zbins}
\end{figure*}

We simultaneously extract E- and B-modes with the quadratic estimator and show the effective-area-weighted six B-mode band powers for two and three redshift bins in Figs.~\ref{fig:signals_BB_2zbins}~and~\ref{fig:signals_BB_3zbins}. The B-mode errors are estimated from the shape noise contribution only, under the assumption that there are no B-modes in the data. This is a very conservative estimate in the sense that it yields the smallest error bars and B-modes not consistent with zero might appear more significant than they are. Following the discussion of Section~\ref{sec:fiducial_B_modes} we corrected the B-modes shown here for the `resetting bias' of the quadratic estimator algorithm discussed in Section~\ref{sec:qe}. 
The corrected B-modes shown in Figs.~\ref{fig:signals_BB_2zbins}~and~\ref{fig:signals_BB_3zbins} can be used as a test for residual systematics in the data, since the cosmological signal is contained entirely in the E-modes in the absence of systematics (Section~\ref{sec:theo}) and the quadratic estimator does not introduce power leakage/mixing either (Appendix~\ref{app:power_leakage}). As we show quantitatively in Section~\ref{sec:results} the corrected B-modes shown here for both redshift bin analyses are indeed consistent with zero. 

\subsection{Cosmological inference}
\label{sec:cosmo_inference}

The cosmological interpretation of the measured (tomographic) band powers $\mathcal{B}_\alpha$ derived in Section~\ref{sec:signal} is carried out in a Bayesian framework. For the estimation of cosmological E-mode and (nuisance) B-mode model parameters $\bmath{p}$ we sample the likelihood 
\begin{equation} \label{eq:shear_lkl}
-2 \ln {\mathcal{L}(\bmath{p})} = \sum_{\alpha, \, \beta} d_\alpha(\bmath{p}) (\mathbfss{C}^{-1})_{\alpha \beta} \, d_\beta(\bmath{p}) \, ,
\end{equation}
where the indices $\alpha$, $\beta$ run over the tomographic bins. The analytical covariance matrix $\mathbfss{C}$ is calculated as outlined in Section~\ref{sec:cov} 
for both E- and B-modes. We note that the assumption of Gaussian band power distributions behind this likelihood is of course only an approximation. However, we show in Fig.~\ref{fig:comparison_E_modes_GRFs} that we recover the means of the bands of interest accurately (see also Section~\ref{sec:fiducial_B_modes}), so that any deviation from Gaussianity only creates an error on the error. Given the current level of uncertainty on the measurements this can be neglected.
  
The components of the data vector are calculated as
\begin{equation}
d_\alpha(\bmath{p}) = \mathcal{B}_\alpha - \langle \mathcal{B}_\alpha(\bmath{p}) \rangle^{\rm model} \, ,
\end{equation}
where the dependence on cosmological parameters enters only in the calculation of the predicted E-mode band powers, $\langle \mathcal{B}^{i}(\ell) \rangle^{\rm model}$ (equations~\ref{eq:conv_window_func} and~\ref{eq:theo_power_spec}). 

For an efficient evaluation of the likelihood 
we employ the nested sampling algorithm {\scriptsize MULTINEST}\footnote{Version 3.8 from \url{http://ccpforge.cse.rl.ac.uk/gf/project/multinest/}} \citep{Feroz2008, Feroz2009, Feroz2013}. Conveniently, its {\scriptsize PYTHON}-wrapper {\scriptsize PYMULTINEST} \citep{Buchner2014} is included in the framework of the cosmological likelihood sampling package {\scriptsize MONTE PYTHON}\footnote{Version 2.2.1 from \url{https://github.com/baudren/montepython_public}} \citep{Audren2013} with which we derive all cosmology-related results in this analysis. 

We note that the likelihood pipeline used here is completely independent from the cosmology pipeline used in \citet{Hildebrandt2016}. However, we verified that it can reproduce the fiducial results from the correlation-function analysis of that study, too. Moreover, we make the likelihood-module written for the {\scriptsize MONTE PYTHON} package publicly available.\footnote{The likelihood module can be downloaded from \url{https://bitbucket.org/fkoehlin/kids450_qe_likelihood_public}}

\subsubsection{Theoretical power spectra}
\label{sec:theo_power_spec}

The calculation of the theoretical shear power spectrum $C_{\mu\nu}(\ell)$ is described in Section~\ref{sec:theo} and summarised by noting that it is the projection of the three-dimensional matter power spectrum $P_\delta$ along the line-of-sight weighted by lensing weight functions $q_\mu$ that take the lensing efficiency of each tomographic bin into account.  

For the calculation of the matter power spectrum $P_\delta(k; \chi)$ in equation~(\ref{eq:theo_power_spec}) we employ the Boltzmann-code {\scriptsize CLASS}\footnote{Version 2.5.0 from \url{https://github.com/lesgourg/class_public}} \citep{Blas2011, Audren2011}. The non-linear corrections are implemented through the {\scriptsize HALOFIT} algorithm including the recalibration by \citet{Takahashi2012}. Additionally, the effects of (massive) neutrinos are also implemented in {\scriptsize CLASS} (\citealt{Class_neutrinos, Bird2012}; see also \citealt{Mead2016} for an alternative non-linear model for massive neutrino cosmologies). Massive neutrinos introduce a redshift- and scale-dependent reduction of power in the matter power spectrum $P_\delta$ mostly on non-linear scales. This reduction of power also propagates into the lensing power spectra $C_{\mu\nu}(\ell)$ smoothed, however, by the lensing weight functions $q_\mu$. In the multipole range considered in this analysis, we expect massive neutrinos to decrease the lensing power spectrum by an almost constant factor. Hence, the effect of massive neutrinos causes a degeneracy with cosmological parameters affecting the normalization of the lensing power spectrum. 

In the following likelihood analysis we always assume a cosmological model with spatially flat geometry and use the same set of key cosmological parameters and priors as in the analysis of \citet{Hildebrandt2016}: ${\Omega_{\rm cdm}h^2, \ \ln (10^{10} A_{\rm s}), \ \Omega_{\rm b}h^2, \ n_{\rm s}, \ h}$, i.e. the amplitude of the primordial power spectrum $A_{\rm s}$, the value $h$ of the Hubble parameter today divided by $100 \, {\rm km/s/Mpc}$, the cold dark matter density $\Omega_{\rm cdm}h^2$, the baryonic matter density $\Omega_{\rm b}h^2$, and the exponent of the primordial power spectrum $n_{\rm s}$. In addition to these we also include the total sum of three degenerate massive neutrinos, $\Sigma m_\nu$.

Moreover, we account for various astrophysical nuisances (Section~\ref{sec:astro_sys}) and always marginalise over the uncertainties of other systematics such as the multiplicative shear calibration bias and redshift distribution $n(z)$ (Section~\ref{sec:other_sys}). 

The employed prior range on $h$ corresponds to the $\pm 5\sigma$ uncertainty centred on the distance-ladder constraint from \citet{Riess2016} of $h = 0.730 \pm 0.018$. Note that the corresponding prior range of $0.64 < h < 0.82$ still includes the preferred value from \citet{Planck2015_CP}. The prior on $\Omega_{\rm b}h^2$ is based on BBN constraints listed in the 2015 update from the Particle Data Group \citep{Olive2014} again adopting a conservative width of $\pm 5\sigma$ such that $0.019 < \Omega_{\rm b}h^2 < 0.026$. 

The cosmic shear power spectrum is mostly sensitive to the two parameters $\Omega_{\rm m}$, the energy density of matter in the Universe today, and $A_{\rm s}$, the amplitude of the primordial power spectrum. These two quantities determine the tilt and the total amplitude of the shear power spectrum, respectively, and are degenerate with each other. In addition to $A_{\rm s}$, the amplitude of the matter power spectrum is also often quantified in terms of $\sigma_8$, the root-mean-square variance in spheres of $8 \, h^{-1} {\rm Mpc}$ on the sky. 

In addition to the parameter combination $\sigma_8 (\Omega_{\rm m}/0.3)^\alpha$ also the quantity $S_8 \equiv \sigma_8 \sqrt{\Omega_{\rm m}/0.3}$ is used in the literature based on the observation that the exponent $\alpha$ is usually close to $0.5$. 

\subsubsection{Astrophysical systematics}
\label{sec:astro_sys}

In order to derive accurate cosmological parameters from the cosmic shear power spectrum measurement it is important to account for a number of astrophysical systematics. 

Feedback from AGN, for example, modifies the matter distribution at small scales (e.g. \citealt{Semboloni2011, Semboloni2013}), resulting in a modification of the dark matter power spectrum at high multipoles. 
Although the full physical description of baryon feedback is not established yet, hydrodynamical simulations offer one route to estimate its effect on the matter power spectrum. In general, the effect is quantified through a bias function with respect to the dark-matter only $P_\delta$ (e.g. \citealt{Semboloni2013, Harnois2015}):
\begin{equation}
b^2(k, z) \equiv \frac{P_{\delta}^{\mathrm{mod}}(k, z)}{P_{\delta}^{\mathrm{ref}}(k, z)} \, ,
\end{equation}
where $P_{\delta}^{\mathrm{mod}}$ and $P_{\delta}^{\mathrm{ref}}$ denote the power spectra with and without baryon feedback, respectively.
 
In this work we make use of the results obtained from the OverWhelmingly Large Simulations (OWLS; \citealt{Schaye2010}, \citealt{vanDaalen2011}) by implementing the fitting formula for baryon feedback from \citet{Harnois2015}:
\begin{equation}
\label{eq:baryon_feedback}
b^2(k, z) = 1 - A_{\mathrm{bary}}[A_z\mathrm{e}^{(B_zx-C_z)^3}-D_zx\mathrm{e}^{E_zx}] \, ,
\end{equation}
where $x=\log_{10}(k/1 \, \mathrm{Mpc}^{-1})$ and the terms $A_z$, $B_z$, $C_z$, $D_z$, and $E_z$ are functions of the scale factor $a = 1/(1+z)$. These terms also depend on the baryonic feedback model and we refer the reader to \citet{Harnois2015} for the specific functional forms and constants. 
Additionally, we introduce a general free amplitude $A_{\rm bary}$ which we will use as a free parameter to marginalise over while fitting for the cosmological parameters.

\citet{Hildebrandt2016} used an alternative description for the baryon feedback model by \citet{Mead2015}, which also includes massive neutrinos on non-linear scales. However, this model is not yet available for {\scriptsize CLASS}. Therefore, we use here the {\scriptsize HALOFIT} algorithm within {\scriptsize CLASS} (including the \citealt{Takahashi2012} recalibration and massive neutrino modelling on non-linear scales by \citealt{Bird2012}) and add the baryon feedback model through equation~(\ref{eq:baryon_feedback}) instead.

Baryon feedback causes a significant reduction of power in the high multipole regime, whereas massive neutrinos lower the amplitude of the shear power spectrum over the scales considered in this analysis by an almost constant value (e.g. fig.~6 in \citealt{Koehlinger2016}, where a similar range of multipoles was used).

Intrinsic alignments (IA) are another important astrophysical systematic, since in general, the observed shear power spectrum $C^\mathrm{tot}$ is a biased tracer of the cosmological convergence power spectrum $C^\mathrm{GG}$:
\begin{equation}
C^\mathrm{tot}_{\mu\nu}(\ell) = C^\mathrm{GG}_{\mu\nu}(\ell) + C^\mathrm{II}_{\mu\nu}(\ell) + C^\mathrm{GI}_{\mu\nu}(\ell) \, ,
\end{equation}
where $C^\mathrm{II}$ is the power spectrum of intrinsic ellipticity correlations between neighbouring galaxies (termed `II') and $C^\mathrm{GI}$ is the power spectrum of correlations between the intrinsic ellipticities of foreground galaxies and the gravitational shear of background galaxies (termed `GI').
We model these effects as in \citet{Hildebrandt2016} and employ the non-linear modification of the tidal alignment model of intrinsic alignments \citep{HirataSeljak2004, BridleKing2007, Joachimi2011}, so that we can write: 
\begin{equation}
\label{eq:cl_II}
C^\mathrm{II}_{\mu\nu}(\ell) = \int_{0}^{\chi_{\rm H}} \diff \chi \, \frac{n_\mu(\chi) n_\nu(\chi) F^2(\chi)}{f_\mathrm{K}^2(\chi)} P_\delta\left(k=\frac{\ell + 0.5}{f_\mathrm{K}(\chi)}; \chi \right) \, ,
\end{equation}
\begin{align} \label{eq:cl_GI}
C^\mathrm{GI}_{\mu\nu}(\ell) = \int_{0}^{\chi_{\rm H}} \diff \chi \, & \frac{q_\nu(\chi) n_\mu(\chi) + q_\mu(\chi) n_\nu(\chi)}{f_\mathrm{K}^2(\chi)} \nonumber \\
 & F(\chi) P_\delta\left(k=\frac{\ell + 0.5}{f_\mathrm{K}(\chi)}; \chi \right) \, , 
\end{align}
with the lensing weight function $q_\mu(\chi)$ defined as in equation~(\ref{eq:lensing_kernel}) and 
\begin{equation}
\label{eq:ia_factor}
F(\chi) = -A_\mathrm{IA} C_1 \rho_\mathrm{crit} \frac{\Omega_{\rm m}}{D_+(\chi)} \, . 
\end{equation}
Here we also introduce a dimensionless amplitude $A_\mathrm{IA}$ which allows us to rescale and vary the fixed normalization $C_1 = 5 \times 10^{-14} \, h^{-2} {\rm M_{\sun}}^{-1} {\rm Mpc}^3$ in the subsequent likelihood analysis. The critical density of the Universe today is denoted as $\rho_\mathrm{crit}$ and $D_+(\chi)$ is the linear growth factor normalised to unity today.
We do not include a redshift or luminosity dependence in the IA modelling as those were found to be negligible by \citet{Joudaki2016}. This model is capable of describing both the well-detected IA signals for elliptical galaxy samples and the null detections reported for samples dominated by disc galaxies.

\subsubsection{Other systematics}
\label{sec:other_sys}

We always marginalise over the uncertainty of the multiplicative shear calibration bias by including the dummy nuisance parameter $m$ as described in Section~\ref{sec:m-correction}. Moreover, we account for the uncertainty in the redshift distribution $n(z)$ (Section~\ref{sec:data_meas}), by drawing in each likelihood evaluation a random realisation of the redshift distribution derived from one of the 1000 bootstrap realisations of the spectroscopic training catalogue. 

The quadratic estimator algorithm also requires a precise and accurate measurement of the noise level in the data (Section~\ref{sec:qe}). 
This can be achieved with a dedicated suite of image simulations aiming at a calibration of the observed ellipticity dispersion. As those simulations were not available at the time this paper has been written, we include a model for excess-noise power in the extracted signals:
\begin{equation}
\label{eq:excess_noise}
 p_\mathrm{noise}(\ell, \, z_i) = A_\mathrm{noise}(z_i) \frac{\ell (\ell + 1)}{2 \pi} \frac{\sigma^2_{\tilde{\gamma}(z_i)}}{n_{\rm eff}(z_i)} \, .
\end{equation}
Here $A_\mathrm{noise}(z_i)$ is a free amplitude that determines the strength of the excess-noise in each redshift auto-correlation power spectrum. 
Since noise contributes equally to E- and B-modes this model is also used in the fitting of the B-mode power spectrum. 
We confirm that including this model is indeed required by the data in the sense of that we find consistent noise amplitudes between E-mode and B-mode only likelihood evaluations (see Appendix~\ref{app:excess_noise} for details).

Finally, in the modelling of the B-modes we account for the `resetting bias' discussed in Section~\ref{sec:fiducial_B_modes}. This is modelled as a power law with two free parameters $A_{{\rm rb}}$ and $\beta_{{\rm rb}}$ (equation~\ref{eq:power_law_B_modes}). In order to marginalise over the uncertainties of these parameters, we draw in each step of the likelihood evaluation random realisations of these parameters from a 2D Gaussian centred on their best-fitting values determined from the GRF fits and we also take their covariance fully into account (Section~\ref{sec:fiducial_B_modes}).  

\section{Results and discussion}
\label{sec:results}

The physical and nuisance parameters discussed in the previous section 
constitute our fiducial model for deriving cosmological parameters to which we refer subsequently as `\lcdm ${+}A_{\rm IA}{+}A_{\rm bary}{+}\Sigma m_\nu{+}{\rm noise}$'. Constraints on all cosmological and nuisance parameters can be found in Table~\ref{tab:cosmo_params} in Appendix~\ref{app:extra_tab} including their prior ranges. In order to highlight parameter degeneracies we show all possible 2D parameter projections for this model in Fig.~\ref{fig:triangle_base_all} in Appendix~\ref{app:extra_fig}. 

\begin{figure*}
	\centering
	\includegraphics[width=156mm]{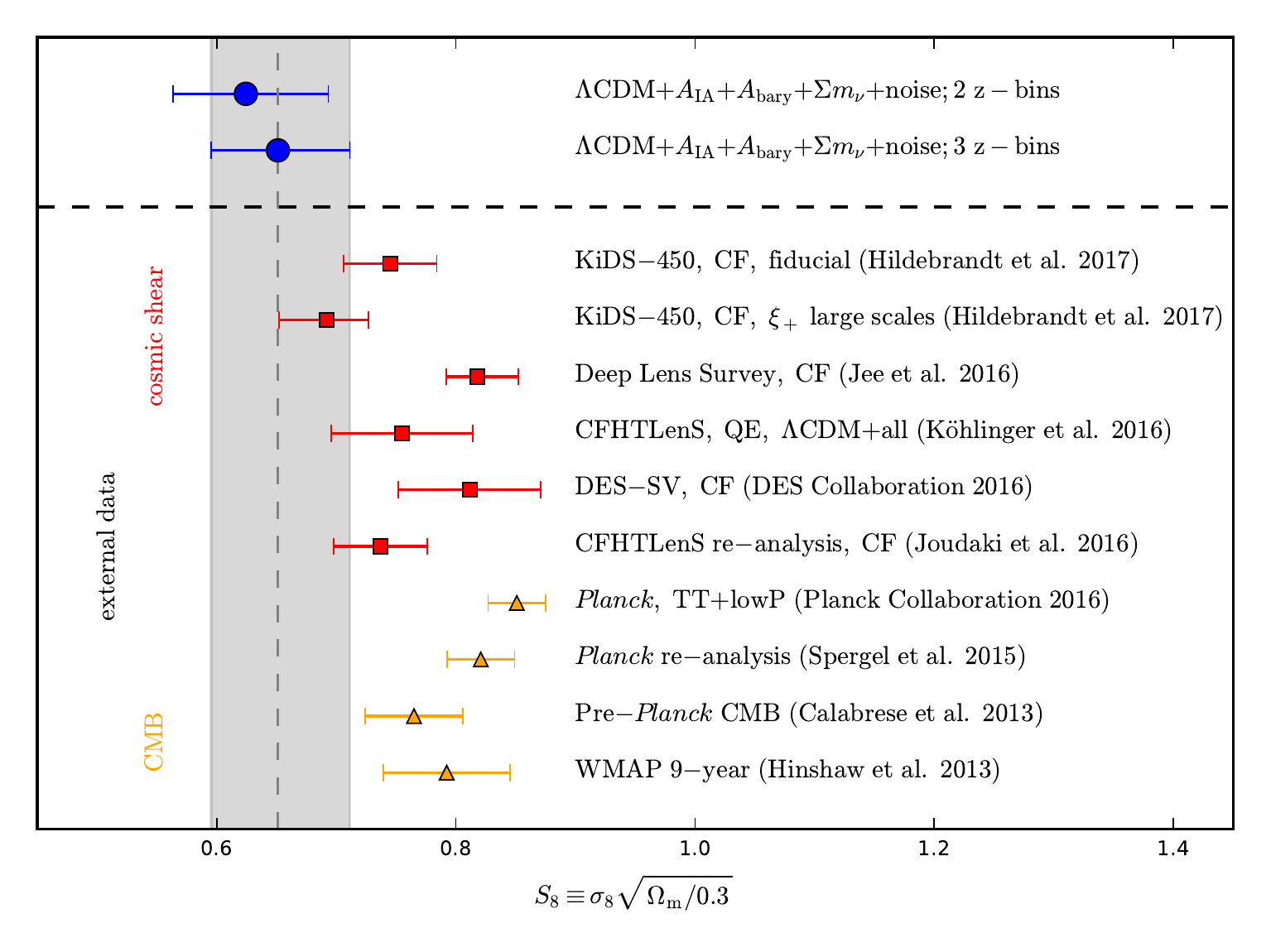}
    \caption{The $1\sigma$-constraints on the parameter combination $S_8 \equiv \sigma_8 \sqrt{\Omega_{\rm m}/0.3}$ for our fiducial model using 2 and 3 redshift bins (Tables~\ref{tab:cosmo_params} and ~\ref{tab:combination}). We compare them to constraints from other cosmic shear and CMB analyses. For cosmic shear analyses we indicate the type of estimator used with `CF' for correlation functions and `QE' for the quadratic estimator.} 
    \label{fig:mymodels_S8}
\end{figure*}

The primary cosmological constraints on $\sigma_8 (\Omega_{\rm m}/0.3)^\alpha$ and $S_8$ are summarised for the 2 z-bin and 3 z-bin analyses in Table~\ref{tab:combination} in Appendix~\ref{app:extra_tab}. The exponent $\alpha$ is derived by fitting the function $\ln \sigma_8(\Omega_{\rm m}) = -\alpha \ln \Omega_{\rm m} + {\rm const.}$ to the likelihood surface in the $\Omega_{\rm m}$--$\sigma_8$ plane. Since indeed $\alpha \approx 0.5$, we compare the $S_8$ values for the 2 z-bin and 3 z-bin analysis in Fig.~\ref{fig:mymodels_S8} to constraints from other cosmic shear analyses and CMB constraints. 

The $S_8$ values we derive for the fiducial models of the 2 z-bin and 3 z-bin analyses are consistent with each other. A comparison of these results with the fiducial results from the correlation function analysis by \citet{Hildebrandt2016} is complicated by the fact that their analysis includes much more information from small scales. At face value our constraints from the quadratic estimator analysis are not consistent with the fiducial result presented in \citet{Hildebrandt2016}, favouring a lower value of $S_8$. As in this work we use larger angular scales compared to the fiducial analysis presented in \citet{Hildebrandt2016}, we also compare our results with the $S_8$ constraints they derived excluding small angular scales from the correlation-function measurements (`$\xi_{+}$ large scales').\footnote{We note that this run did not marginalise over uncertainties in the redshift distribution nor over baryons, only marginalisation over IA was included.} The lower $S_8$ value reported by \citet{Hildebrandt2016} is in broad agreement with the result presented in this work.
The same trend of a lower $S_8$ value for more conservative small-scale cuts is also found by \citet{Joudaki2016b}, who considered a more conservative large-scale case\footnote{They considered only two bins for $\xi_+$ at 25 arcmin and 51 arcmin, and one bin in $\xi_-$ at 210 arcmin.} in their extended cosmological analysis of the KiDS-450 correlation function results. 
We remind the reader though to be cautious when quantifying tension between datasets based on parameter projections of multidimensional likelihoods (see appendix~A in \citealt{MacCrann2015}).

In Fig.~\ref{fig:2d_projection}(a) we show constraints in the $S_8$ versus $\Omega_{\rm m}$ plane and note that the tension observed in the one-dimensional projection of $S_8$ between results from this analysis and the fiducial correlation-function analysis from \citet{Hildebrandt2016} is weaker resulting in a large overlap of the 68 and 95 per cent credibility intervals. As expected from the consistency of the $S_8$ values when comparing to their large angular scale analysis (`KiDS-450, CF, $\xi_{+}$ large scales'), the 68 and 95 per cent credibility contours show both a substantial overlap in the two-dimensional parameter projection as shown in Fig.~\ref{fig:2d_projection}(b). The tension between the results derived here and constraints from \citet[`TT+lowP']{Planck2015_CP}, however, is significant since the 68 and 95 per cent credibility intervals do not overlap in this projection.  

An accurate estimate of the statistical significance of the differences between the quadratic estimator and the correlation function analyses applied to the same dataset is complicated as it requires an \"{u}ber-covariance of the estimators. This comparison, although interesting, is beyond the scope of this paper.

Our model in both redshift bin analyses is also consistent with previous results from CFHTLenS, where we compare in particular to a correlation-function re-analysis employing seven tomographic bins and marginalization over key astrophysical systematics from \citet{Joudaki2016}. In addition to that, we show results from our previous quadratic estimator analysis of CFHTLenS \citep{Koehlinger2016}, which employed two tomographic bins at higher redshift compared to the redshift bins used here. The label `\lcdm +all' used in that study refers to an extension of a flat \lcdm \ base model with a free total neutrino mass and marginalization over baryon feedback, but does not take intrinsic alignments into account. The errors are comparable to the errors in this study, since CFHTLenS and KiDS-450 have comparable statistical power. 
Our results in this parameter projection disagree mildly with the result from the DES science verification (SV) correlation-function analysis (\citealt{DES_cosmo2015}, `Fiducial DES SV cosmic shear') by $1.9\sigma$ (3 z-bins) / $2.1\sigma$ (2 z-bins). 

Also interesting is the comparison of our results to CMB constraints including pre-\textit{Planck} \citep{Hinshaw2013, Calabrese2013} and \textit{Planck} \citep{Planck2015_CP, Spergel2015} data.
We find them to be most distinctively in tension with the results from \citet{Planck2015_CP} at $3.2\sigma$ (3 z-bins) / $3.3\sigma$ (2 z-bins) which cannot be explained by projecting a multidimensional likelihood into this 1D parameter space alone. 

\begin{figure*}
	\centering
	\begin{subfigure}{0.49\textwidth}
       \centering
       \includegraphics[width=\textwidth]{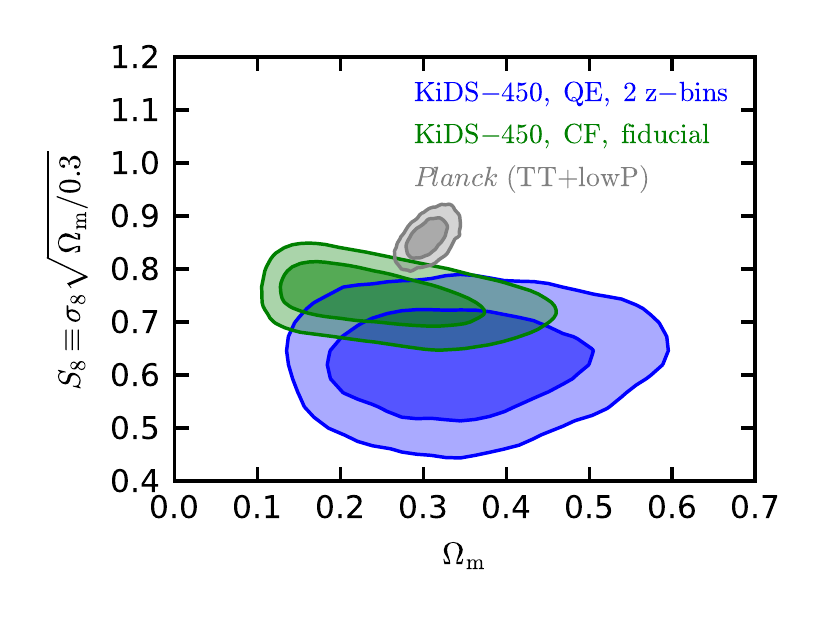}
       \caption{}
    \end{subfigure}
    \begin{subfigure}{0.49\textwidth}
        \centering
        \includegraphics[width=\textwidth]{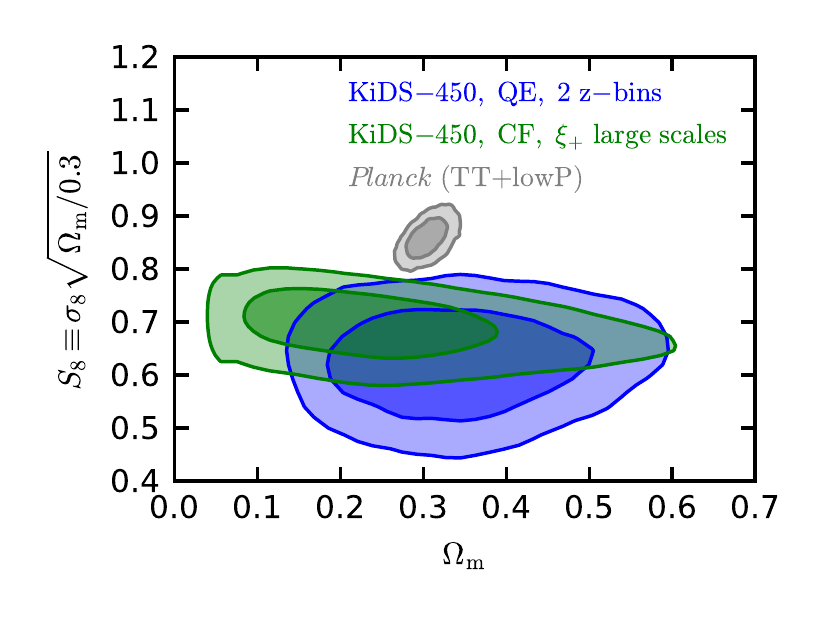}
        \caption{}
    \end{subfigure}
    \caption{(a): Projection of cosmological constraints in the $S_8$ versus $\Omega_{\rm m}$ plane from the KiDS-450 analysis presented here (`KiDS-450, QE, 2 z-bins') and the fiducial correlation-function analysis by \citet[`KiDS-450, CF, fiducial']{Hildebrandt2016}. For comparison we also show contours from \citet[`TT+lowP']{Planck2015_CP}. The inner contours correspond to the 68 per cent credibility interval and the outer ones to the 95 per cent credibility interval. Note that the contours are smoothed with a Gaussian for illustrative purposes only. We chose to present the weaker constraints from the 2 z-bin analysis because that analysis yields the largest tension with respect to the other results. The corresponding figures for the 3 z-bin analysis are presented in Appendix~\ref{app:extra_fig}. (b): The same as in (a) but comparing to the `$\xi_{+}$ large scales' correlation-function analysis from \citet{Hildebrandt2016}.}
	\label{fig:2d_projection}
\end{figure*}

\subsection{Neutrino masses}
\label{sec:neutrinos}

We also derive an upper bound on the total mass for three degenerate massive neutrinos and find $\Sigma m_\nu < 3.3 \, {\rm eV}$ (3 z-bins) / $\Sigma m_\nu < 4.5 \, {\rm eV}$ (2 z-bins) at 95 per cent credibility from lensing alone. 
\citet{Joudaki2016b} also derive a neutrino constraint based on the 4 z-bin correlation-function analysis of the KiDS-450 data \citep{Hildebrandt2016} and find $\Sigma m_\nu < 4.0 \, {\rm eV}$ and $\Sigma m_\nu < 3.0 \, {\rm eV}$ at 95 per cent credibility, the latter depending on the choice of the $H_0$ prior. We note that \citet{Joudaki2016b} use a different implementation of massive neutrinos through {\scriptsize HMCODE} \citep{Mead2016}, whereas the massive neutrino implementation used in the pipeline here is the one from {\scriptsize CLASS} \citep{Class_neutrinos, Bird2012}. We note further that both massive neutrino calibrations are most accurate only for total neutrino masses $\Sigma m_\nu \lesssim 1 \, {\rm eV}$.
So far, these lensing-only constraints on the upper bound of the total mass of neutrinos are still weaker than non-lensing constraints as found by \citet[`TT+lowP']{Planck2015_CP}, who report $\Sigma m_\nu < 0.72 \, {\rm eV}$ at 95 per cent confidence. Combining the \textit{Planck} CMB results with measurements of the Ly$\, \alpha$ power spectrum and BAO measurements yields the very stringent upper limit of $\Sigma m_\nu < 0.14 \, {\rm eV}$ at 95 per cent confidence \citep{Palanque2015}. 

\subsection{Error budget}
\label{sec:errors}

Comparing the error bars between our quadratic-estimator 2 z-bin and 3 z-bin analyses and the 4 z-bin correlation-function analysis by \citet{Hildebrandt2016} we note that the error bars of the correlation function results are smaller by more than a factor of two in comparison to our power spectrum analysis (while marginalizing over a comparable set of nuisance parameters except for the excess-noise). Comparing the error bars of the 2 z-bin analysis versus the 3 z-bin analysis we find that using more redshift bins in a tomographic analysis improves the constraints despite lowering the SNR per individual redshift bin. However, this effect cannot explain the larger error bars in this power spectrum analysis and one should also consider the information from smaller scales that has entered the correlation-function analysis (see Appendix~\ref{sec:theta_comparison} or fig.~4 in \citealt{Kilbinger2017}). 

\begin{figure*}
    	\centering
		\begin{subfigure}{0.49\textwidth}
			\centering			
			\includegraphics[width=\textwidth]{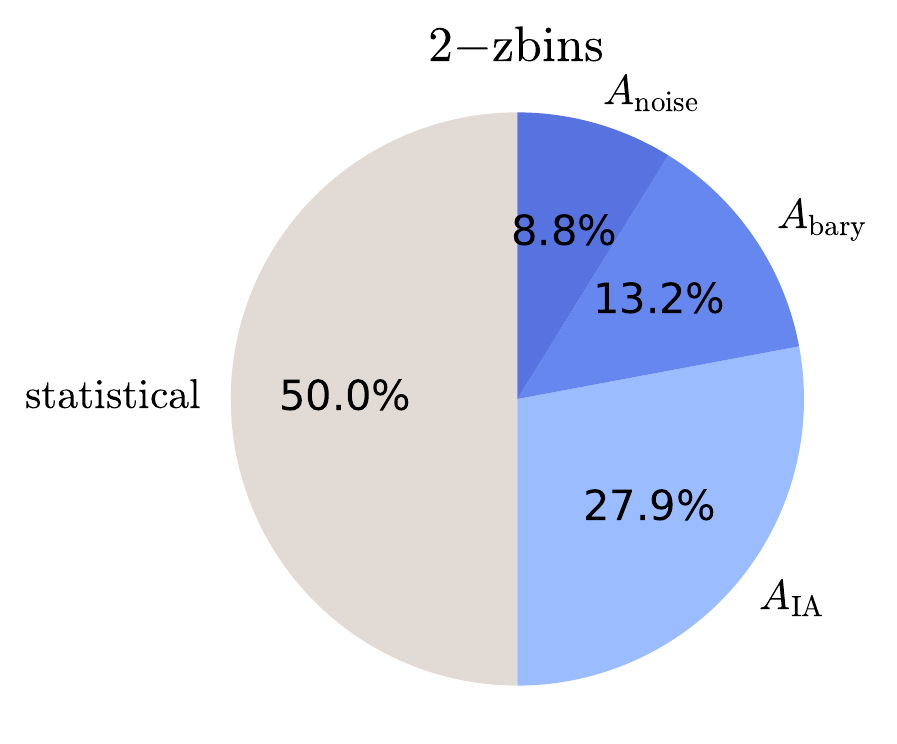}
			\caption{}
			\label{fig:errors_2zbins}
		\end{subfigure}
        \begin{subfigure}{0.49\textwidth}
           	\centering
			\includegraphics[width=\textwidth]{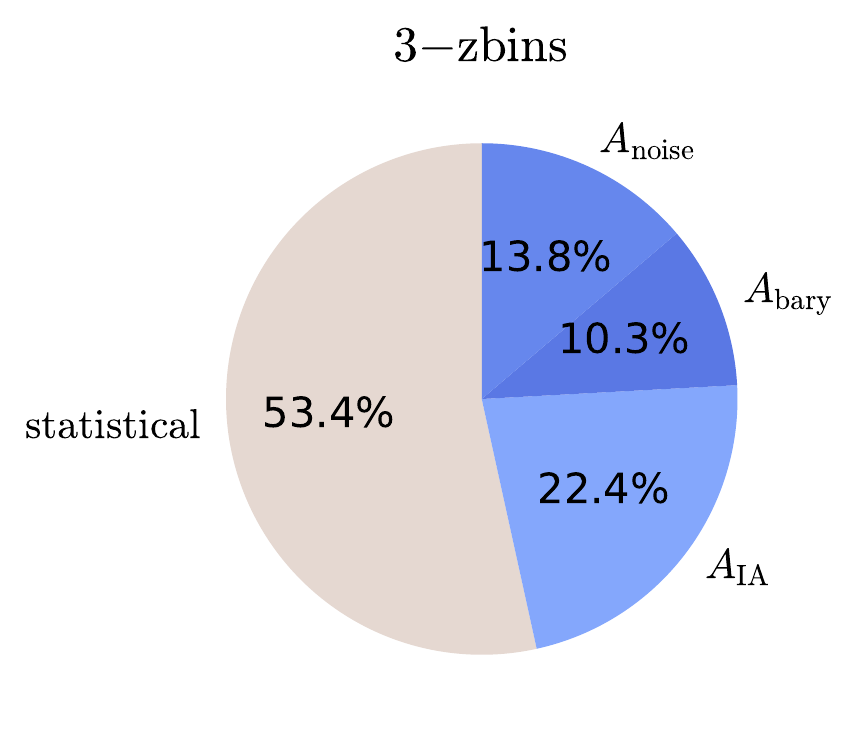}
			\caption{}
			\label{fig:errors_3zbins}
		\end{subfigure}
\caption{(a) Error budget for the parameter combination $S_8$ in the 2 z-bin analysis for the fiducial model shown in Fig.~\ref{fig:mymodels_S8} and described in Section~\ref{sec:cosmo_inference}. The statistical error already includes marginalisation over the source redshift distribution, shear calibration, and resetting bias. The uncertainty due to marginalising over baryon feedback and intrinsic alignments is denoted as $A_{\rm bary}$ and $A_{\rm IA}$, respectively. The uncertainty due to marginalising over residual excess-noise is labelled by $A_{\rm noise}$. The radius of the pie chart is set to the ratio of the total error of the 2 z-bin analysis over the total error of the 3 z-bin analysis. (b) The same as (a) but for the 3 z-bin analysis. The radius of the pie chart is set to one.}
	\label{fig:error_budget}
\end{figure*}

In order to give the reader a feeling for the relative contribution of the different sources of uncertainty to our final error budget on the parameter $S_8$, we show in Figs.~\ref{fig:error_budget}(a)~and~(b) a detailed error budget for the 2 z-bin and 3 z-bin analyses. The dominant part of the uncertainty is already set by the statistical error which also accounts for the marginalization over the shear calibration and redshift distribution uncertainty. Other sources of uncertainty and their relative impact differs between the 2 z-bin and 3 z-bin analysis due to their different redshift sensitivity. 
Adding more tomographic bins to the analysis decreases the uncertainty due to marginalizing over the intrinsic alignment modelling, as expected.
However, for the quadratic estimator adding more and more redshift bins becomes impractical due to the strong dependence of the matrix dimensionality on the number of redshift bins and hence runtime (Section~\ref{sec:qe}).

Improving upon the prior for the baryon feedback model is also worth pursuing since the uncertainty due to $A_{\rm bary}$ contributes about $10$ per cent to the total uncertainty.

A further limitation for high-precision constraints with the quadratic estimator is the requirement to marginalise over the excess-noise power model. 
Although this uncertainty contributes only $8.8$ per cent to the total uncertainty in the 2 z-bin analysis it is strongly dependent on redshift and its contribution rises to $13.8$ per cent in the 3 z-bin analysis. The 3 z-bin analysis is more affected because the SNR per bin is lower in this case. However, the 3 z-bin analysis still yields a smaller total uncertainty on $S_8$, so it is worthwhile to investigate further mitigation strategies for the excess-noise power contribution. Equivalently a stringent quantification of the uncertainty of the root-mean-square ellipticity dispersion would allow us to use more informative priors for the parameters of the excess-noise model. Moreover, we note that correlation-function measurements are also affected by excess-noise through their covariance matrix in which a biased estimate of the shear dispersion enters (although this can both increase or decrease the errors depending on the bias).

Finally, we note that the calibration model for the fiducial B-modes (Section~\ref{sec:fiducial_B_modes}) yields B-modes consistent with zero in the four bands considered for deriving cosmological constraints. This is shown in Figs.~\ref{fig:signals_BB_2zbins}~and~\ref{fig:signals_BB_3zbins} in which the `resetting bias' model was subtracted off the extracted B-modes. We assess the consistency of the corrected B-modes with zero more quantitatively via a $\chi^2$-goodness-of-fit measure and find: $\chi^2_{\rm red} = 1.10$ for $23$ degrees of freedom in the 3 z-bin analysis and $\chi^2_{\rm red} = 0.58$ for $11$ degrees of freedom in the 2 z-bin analysis. Hence, there is no significant B-mode contamination in the data for the scales used in this analysis.
This implies that the small residual B-mode contamination on small angular scales observed by \citet{Hildebrandt2016} is indeed most likely caused by some unknown systematic affecting the high multipoles that we do not include in our analysis presented here.

\section{Conclusions}
\label{sec:conclusions}

In this study we applied the quadratic estimator to shear data from KiDS-450 in two and three redshift bins over the range $0.10 < z_{\rm B} \leq 0.90$ and extracted the band powers of the auto-correlation and cross-correlation shear power spectra for multipoles in the range $76 \leq \ell \leq 1310$. The covariance matrix is based on an analytical calculation which is then convolved with the full band window matrix. We interpret our measurements in a Bayesian framework and we derive cosmological parameters after marginalizing the posterior distribution over a free total neutrino mass, physical nuisances such as intrinsic alignments and baryon feedback, and nuisance parameters for excess noise. The model also includes a marginalization over the small uncertainties of the shear calibration and accounts for the uncertainty of the redshift distributions. 

We find $S_8 = 0.651 \pm 0.058$ (3 z-bins) /  $S_8 = 0.624 \pm 0.065$ (2 z-bins), which is in tension with the value from \textit{Planck} at $3.2\sigma$ (3 z-bins) / $3.3\sigma$ (2 z-bins). This supports the result from the fiducial KiDS-450 correlation-function analysis in four tomographic bins by \citet{Hildebrandt2016} with higher significance despite increased error bars by almost a factor of two in comparison to the correlation-function analysis. Moreover, the fact that this study uses fewer of the very non-linear scales in comparison to \citet{Hildebrandt2016} also refute the idea that insufficient modelling of these non-linear scales is a possible explanation for the discrepancy with \textit{Planck}. We emphasize that the estimator, signal extraction and cosmological inference pipelines are independent from the pipelines used in \citet{Hildebrandt2016}; both studies only have the shear catalogues in common. Hence, this study presents an independent cross-check of the previously reported results with respect to the data pipelines.

Finally, we summarise the properties of the quadratic estimator with respect to the steadily increasing amount of data from current and future surveys: although the quadratic estimator is an intrinsically slow matrix algorithm, dealing with shear data of the order of (several) $1000 \, \deg^2$ is in principle still feasible. However, increasing the number of tomographic bins and multipole bins will require major revisions of the code. Porting it, for example, to graphical processing units specifically designed for matrix operations might be the most straightforward solution to this problem. For that purpose we make our code implementation available to the community.\footnote{The quadratic estimator source code can be downloaded from \url{https://bitbucket.org/fkoehlin/qe_public}}
Following a hybrid-approach, also taken for the measurement of CMB power spectra, might alleviate the runtime problem: there the quadratic estimator analysis is limited to include only the largest scales / lowest multipoles and higher multipole bands are measured with intrinsically faster pseudo-$C(\ell)$ methods which are usually less accurate on the largest scales \citep{Asgari2016}.
  
It is also important to realise that a shear calibration produced with a correlation-function analysis in mind might not be optimal for other estimators. In particular, the quadratic estimator can easily account for the effect of a global additive shear bias whose calibration for correlation functions requires significant resources and efforts. However, the noise level in the data must be known very precisely and accurately in order to extract unbiased shear power spectra with the quadratic estimator, whereas in correlation-function measurements the noise level enters only through the covariance. Although the bias can be modelled and mitigated for the quadratic-estimator analysis its mitigation is a major source of uncertainty, especially for an increasing number of tomographic bins. 

\section*{Acknowledgements}
We thank the anonymous referee, Marika Asgari, and Alexander Mead for very helpful comments and discussions that improved the manuscript and its presentation.\\
We acknowledge support from a de Sitter Fellowship of the Netherlands Organization for Scientific Research (NWO) under grant number 022.003.013, the World Premier International Research Center Initiative (WPI), MEXT, Japan, the European Research Council under FP7 grant number 279396 and grant number 647112, an STFC Ernest Rutherford Fellowship and Grant, grant references ST/J004421/1 and ST/L00285X/1, NWO through grants 614.001.103, an Emmy Noether grant (No. Hi 1495/2-1) and in the framework of the TR33 `The Dark Universe' of the Deutsche Forschungsgemeinschaft, the Australian Research Council Centre of Excellence for All-sky Astrophysics (CAASTRO), through project number CE110001020, the People Programme (Marie Curie Actions) of the European Union's Seventh Framework Programme (FP7/2007-2013) under REA grant agreement number 627288, and by the Alexander von Humboldt Foundation. \\
We are grateful to the Lorentz Center for hosting our workshops.\\
Based on data products from observations made with ESO Telescopes at the La Silla Paranal Observatory under programme IDs 177.A-3016, 177.A-3017 and 177.A-3018, and on data products produced by Target/OmegaCEN, INAF-OACN, INAF-OAPD and the KiDS production team, on behalf of the KiDS consortium.\\
\small{ \textit{Author Contributions:} All authors contributed to the development and writing of this paper. The authorship list is given in
three groups: the lead authors (FK, MV, BJ, HH, EvU), followed
by two alphabetical groups. The first alphabetical group (HHi) includes
those who are key contributors to both the scientific analysis and
the data products. The second group covers those who have either
made a significant contribution to the data products, or to the
scientific analysis.}



\bibliographystyle{mnras}
\bibliography{bibliography}


\appendix
\section{Updated derivation of the window function matrix}
\label{app:BWM_update}

In Section~\ref{sec:qe} we noted that the notation of the window function matrix $\mathbfss{W}$ in equation~(\ref{eq:conv_window_func}) has changed with respect to the one given in \citet{Koehlinger2016}. This is necessary, because in order to propagate the properties of the quadratic estimator into the analytical covariance (Section~\ref{sec:cov}), the full band window matrix with all possible cross-terms is required. Hence, we give the updated notation below.

The elements of the window function matrix can be derived as (cf. \citealt{Lin2012})
\begin{equation} \label{eq:window}
W_{A(\zeta\vartheta)}(\ell) = \sum_B \tfrac{1}{2}(\mathbfss{F}^{-1})_{AB} T_{B(\zeta\vartheta)}(\ell) \, ,
\end{equation} 
where $\mathbfss{F}^{-1}$ denotes the inverse of the Fisher matrix (equation~12 in \citealt{Koehlinger2016}). The full index notation for all matrices and tensors used in the quadratic-estimator algorithm can be found in appendix~A of \citet{Koehlinger2016}. The trace matrix $\mathbfss{T}$ is defined as
\begin{equation} \label{eq:trace}
T_{A(\zeta\vartheta)}(\ell) = \tr[\mathbfss{C}^{-1}\mathbfss{D}_{A}\mathbfss{C}^{-1}\mathbfss{D}_{\zeta\vartheta}(\ell)] \, .
\end{equation}
The derivative $\mathbfss{D}_{\zeta\vartheta}(\ell)$ denotes the derivative of the full covariance $\mathbfss{C}$ with respect to the power at a single multipole $\ell$ (per band type $\vartheta$ and unique redshift correlation $\zeta$ and is derived as:
\begin{align}
\label{eq:deriv_BWM}	 
	\frac{\partial C_{(\mu\nu)(ab)(ij)}}{\partial \mathcal{B}_{\zeta\vartheta}(\ell)} &= \frac{M_{\zeta(\mu\nu)}}{2(\ell+1)} [ w_0(\ell)I^{\vartheta}_{(ab)(ij)}\\
	& \hphantom{{} = } + \tfrac{1}{2} \, w_4(\ell) Q^{\vartheta}_{(ab)(ij)} ] \, \nonumber \\ 
	&\equiv D_{(\mu\nu)(ab)(ij)(\zeta)(\vartheta)}(\ell) \equiv \mathbfss{D}_{\zeta\vartheta}(\ell) \, ,
\end{align}
where we have used that
\begin{align}
C^{{\rm sig}}_{(\mu\nu)(ab)(ij)} &= \sum_{\zeta, \vartheta, \ell} \mathcal{B}_{\zeta\vartheta}(\ell) \frac{M_{\zeta(\mu\nu)}}{2(\ell+1)} [ w_0(\ell)I^{\vartheta}_{(ab)(ij)}\\
& \hphantom{{} = } + \tfrac{1}{2} \, w_4(\ell) Q^{\vartheta}_{(ab)(ij)} ] \, \nonumber \, .
\end{align}

\section{Additional tests}
\label{app:extra_tests}

We present here additional tests performed in order to firstly show that the contribution of power leakage/mixing (e.g. due to the survey mask) is negligible for the band powers extracted with the quadratic estimator. In addition we show that power leakage/mixing is not the source of the fiducial B-modes discussed in Section~\ref{sec:fiducial_B_modes}. 

Secondly, we verify that the excess-noise model (Section~\ref{sec:other_sys}) is indeed required in the interpretation of the band power measurements. 

\subsection*{Power leakage/mixing}
\label{app:power_leakage}

\begin{figure*}
	\centering
	\includegraphics[width=180mm]{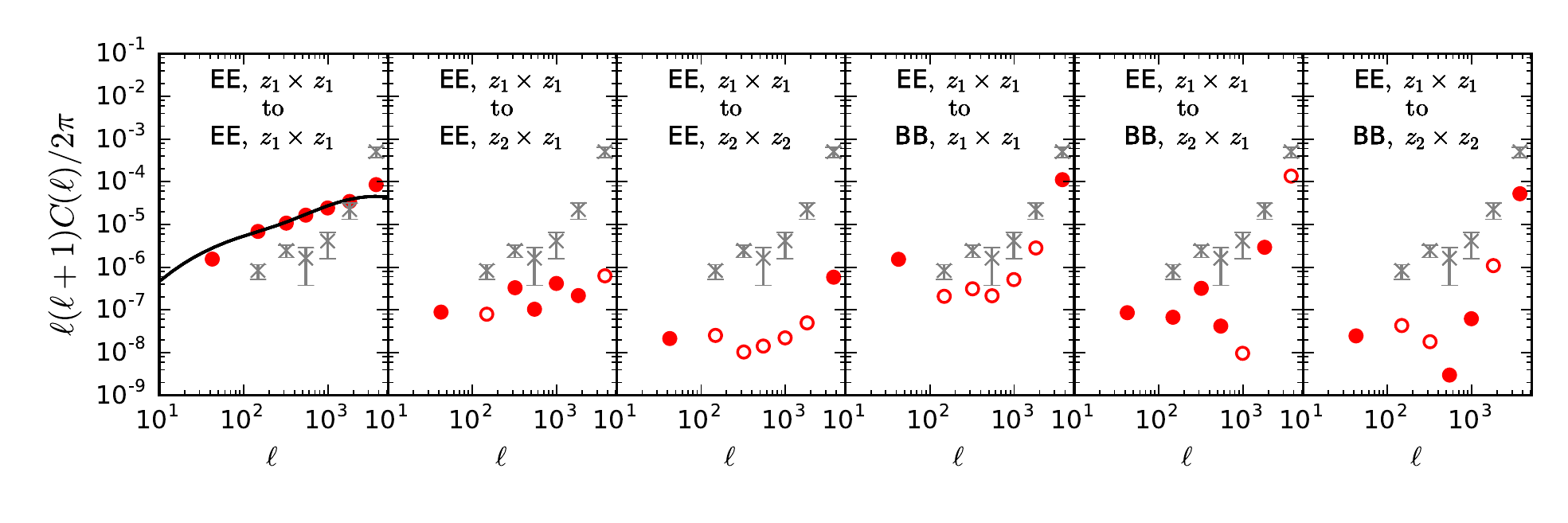}
	\caption{We show that power leakage from E- to B-modes is negligible using 50 GRF realisations of the CFHTLenS W2 field in two tomographic bins. The extracted B-mode band powers (grey crosses with error bars) for the low redshift auto-correlation ($z_1 \times \, z_1$) are compared to the convolution (red points) of the input E-mode signal (`WMAP9'-like cosmology; solid line) with the corresponding band window functions of all possible cross-terms (from left to right; e.g. EE, $z_1 \times \, z_1$ to BB, $z_2 \times \, z_2$). If power leakage/mixing were indeed the cause for these fiducial B-modes (Section~\ref{sec:fiducial_B_modes}), we would expect the convolved E-mode power to match the extracted B-modes, especially in the redshift auto-correlation panel (`EE $z_1 \times \, z_1$ to BB $z_1 \times \, z_1$').}
	\label{fig:BWM_EE_z1z1_conv}
\end{figure*}

\begin{figure*}
	\centering
	\includegraphics[width=180mm]{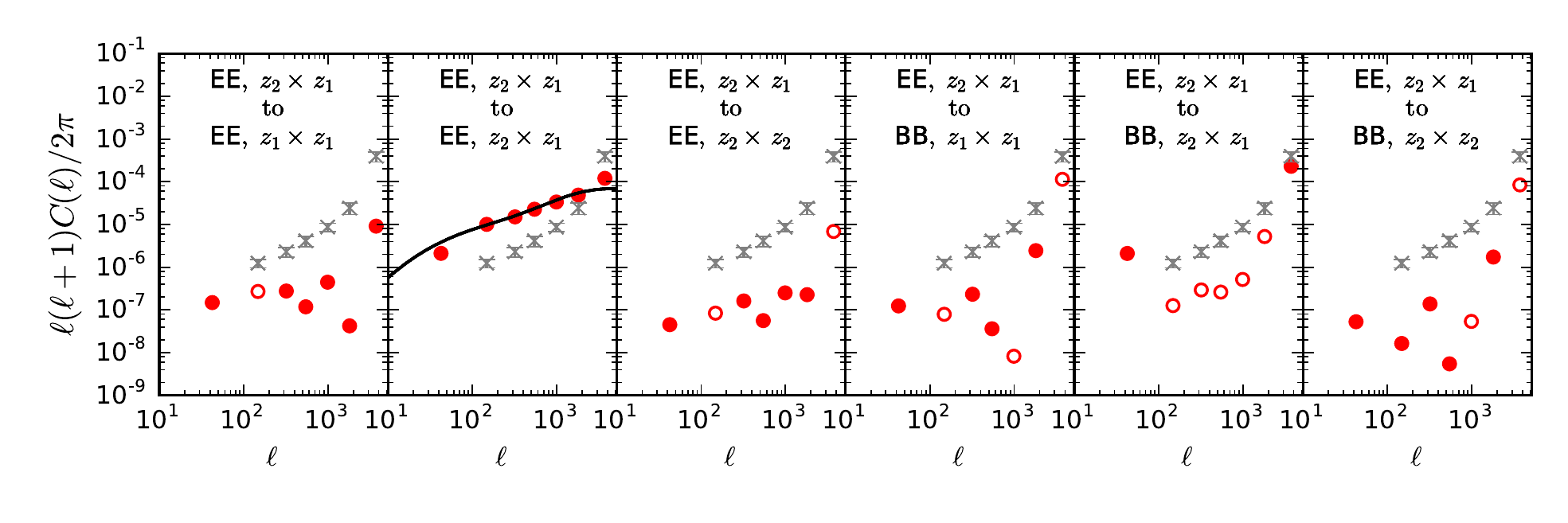}
	\caption{The same as Fig.~\ref{fig:BWM_EE_z1z1_conv} but for the extracted B-modes of the redshift cross-correlation ($z_2 \times \, z_1$). If power leakage/mixing were indeed the cause for these fiducial B-modes (Section~\ref{sec:fiducial_B_modes}), we would expect the convolved E-mode power to match the extracted B-modes, especially in the redshift cross-correlation panel (`EE $z_2 \times \, z_1$ to BB $z_2 \times \, z_1$').}
	\label{fig:BWM_EE_z2z1_conv}
\end{figure*}

\begin{figure*}
	\centering
	\includegraphics[width=180mm]{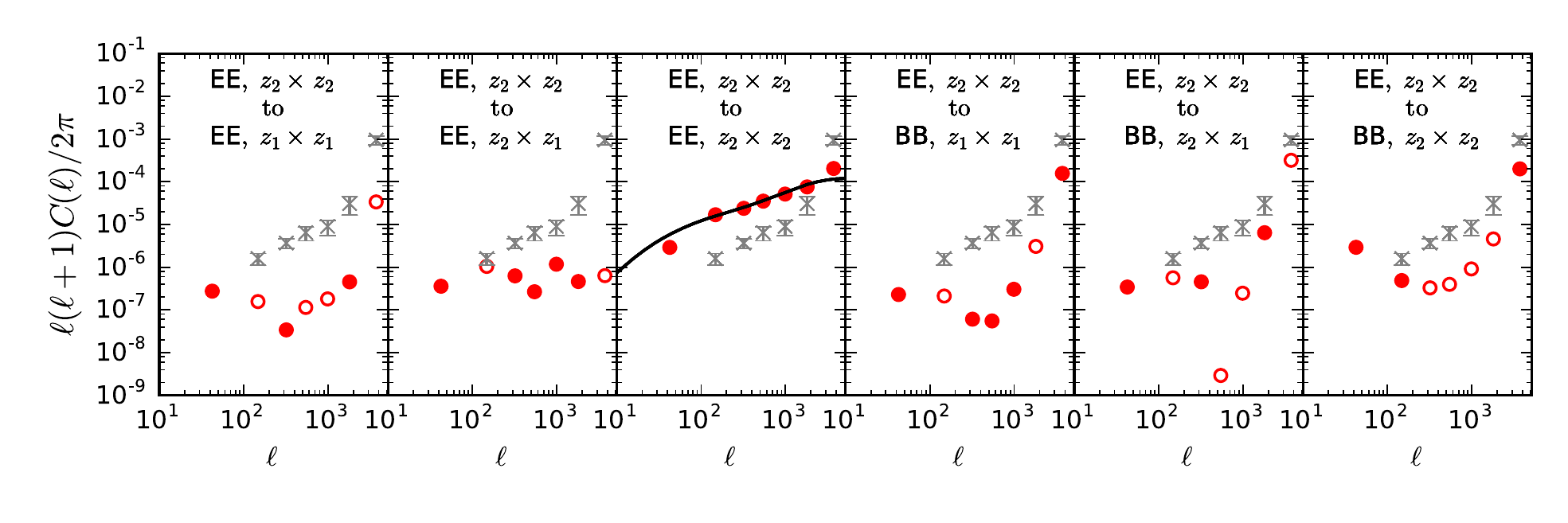}
	\caption{The same as Fig.~\ref{fig:BWM_EE_z1z1_conv} but for the extracted B-modes of the high-redshift auto-correlation ($z_2 \times \, z_2$). If power leakage/mixing were indeed the cause for these fiducial B-modes (Section~\ref{sec:fiducial_B_modes}), we would expect the convolved E-mode power to match the extracted B-modes, especially in the redshift cross-correlation panel (`EE $z_2 \times \, z_2$ to BB $z_2 \times \, z_2$').}
	\label{fig:BWM_EE_z2z2_conv}
\end{figure*}

\citet{Lin2012} showed that power leakage/mixing from E- to B-modes is negligible for the quadratic estimator. However, power leakage/mixing could also be a potential source of the fiducial B-modes discussed in Section~\ref{sec:fiducial_B_modes}. In order to verify that this is also negligible here, we use once more a suite of 50 GRF realisations of the CFHTLenS W2 field in two broad redshift bins (i.e. $z_1{:} \ 0.50 < z_{\rm B} \leq 0.85$ and $z_2{:} \ 0.85 < z_{\rm B} \leq 1.30$; see \citealt{Koehlinger2016} for details) and show the extracted B-mode band powers (grey crosses with error bars) for the low-redshift auto-correlation ($z_1 \times \, z_1$) in Fig.~\ref{fig:BWM_EE_z1z1_conv}, the high-redshift auto-correlation ($z_2 \times \, z_2$) in Fig.~\ref{fig:BWM_EE_z2z2_conv}, and their cross-correlation ($z_2 \times \, z_1$) in Fig.~\ref{fig:BWM_EE_z2z1_conv}. In each figure from left to right panels, these extracted B-modes are compared to the convolution (red points) of the input E-mode signal (`WMAP9'-like cosmology; solid line) with the corresponding band window functions of all possible cross-terms (e.g. EE, $z_1 \times \, z_1$ to BB, $z_2 \times \, z_2$). If power leakage/mixing were indeed the cause for these fiducial B-modes, we would expect the convolved E-mode power to match the extracted B-modes, especially in the redshift auto-correlation panels (`EE $z_1 \times \, z_1$ to BB $z_1 \times \, z_1$') of Figs.~\ref{fig:BWM_EE_z1z1_conv}~and~\ref{fig:BWM_EE_z2z2_conv} and cross-correlation panel of Fig.~\ref{fig:BWM_EE_z2z1_conv}.       

\subsection*{Consistency of the excess-noise model}
\label{app:excess_noise}

For cross-checking whether the excess-noise power model defined in Section~\ref{sec:theo_power_spec} is required by the data, we evaluate the E-mode signal only using a minimal cosmological model (including intrinsic alignments) with the excess-noise model. The recovered amplitudes of the excess-noise model are then compared to evaluations of the B-mode signal only. We find that the noise amplitudes per redshift auto-correlation agree within their error bars across the E-only and B-only inferences (Table~\ref{tab:noise_amplitudes}). This indicates that excess-noise power is contributing equally to both E- and B-modes as expected from theory (Section~\ref{sec:theo}). We note that the negative sign for $A_{\rm noise}(z_3)$ in Table~\ref{tab:noise_amplitudes} implies that the noise is overestimated in that redshift bin, hence the excess-noise model compensates for that by subtracting off the noise component.

\begin{table}
	\caption{Noise amplitudes for separately evaluated E-mode and B-mode signals.}
	\label{tab:noise_amplitudes}
	   \begin{center}
		\begin{tabular}{ l c c }
			\toprule 
			& E-modes only& B-modes only\\			
			\midrule 
			\textbf{2 z-bins:}\\
			$A_{\rm noise}(z_1)$& $-0.014_{-0.017}^{+0.016}$& $-0.015_{-0.015}^{+0.013}$\\[1mm]
			$A_{\rm noise}(z_2)$& $-0.013_{-0.019}^{+0.018}$& $0.023_{-0.013}^{+0.013}$\\[1mm]
			\textbf{3 z-bins:}\\
			$A_{\rm noise}(z_1)$& $-0.016_{-0.018}^{+0.015}$& $-0.009_{-0.010}^{+0.015}$\\[1mm]
			$A_{\rm noise}(z_2)$& $0.035_{-0.016}^{+0.015}$& $0.024_{-0.013}^{+0.012}$\\[1mm]
			$A_{\rm noise}(z_3)$& $0.009_{-0.015}^{+0.015}$& $0.016_{-0.013}^{+0.012}$\\[1mm]
			\bottomrule 
		\end{tabular}
	\end{center}
\end{table}

\section{Comparison to correlation functions}
\label{sec:theta_comparison}

Most cosmic shear studies to date employ real-space correlation functions (e.g. \citealt{Heymans2013, Becker2015, Hildebrandt2016}) because they are conceptually easy and fast to compute. 

In contrast to direct power spectrum estimates, correlation functions measured at a given angular separation sum up contributions over a wide range of multipoles. Due to this mode mixing it is non-trivial to compare angular scales to multipole ranges, as well as to cleanly separate linear and non-linear scales.

A direct power spectrum estimation, however, requires an accurate and precise estimation of the noise level in the data, whereas a measurement of that is not required in the signal extraction step for correlation functions. In this case the accurate noise level estimation enters only in the cosmological likelihood evaluation through the covariance matrix.

As an example, here we qualitatively compare correlation-function measurements based on the angular scales presented in \citet{Hildebrandt2016} to the direct power-spectrum measurements employing the quadratic estimator. For that purpose we calculate a fiducial shear power spectrum (equation~\ref{eq:theo_power_spec}) employing a \textit{Planck} cosmology \citep{Planck2015_CP} and the redshift distributions derived for the 2 z-bin analysis (Table~\ref{tab:n_eff}).
 
\begin{figure}
	\includegraphics[width=\columnwidth]{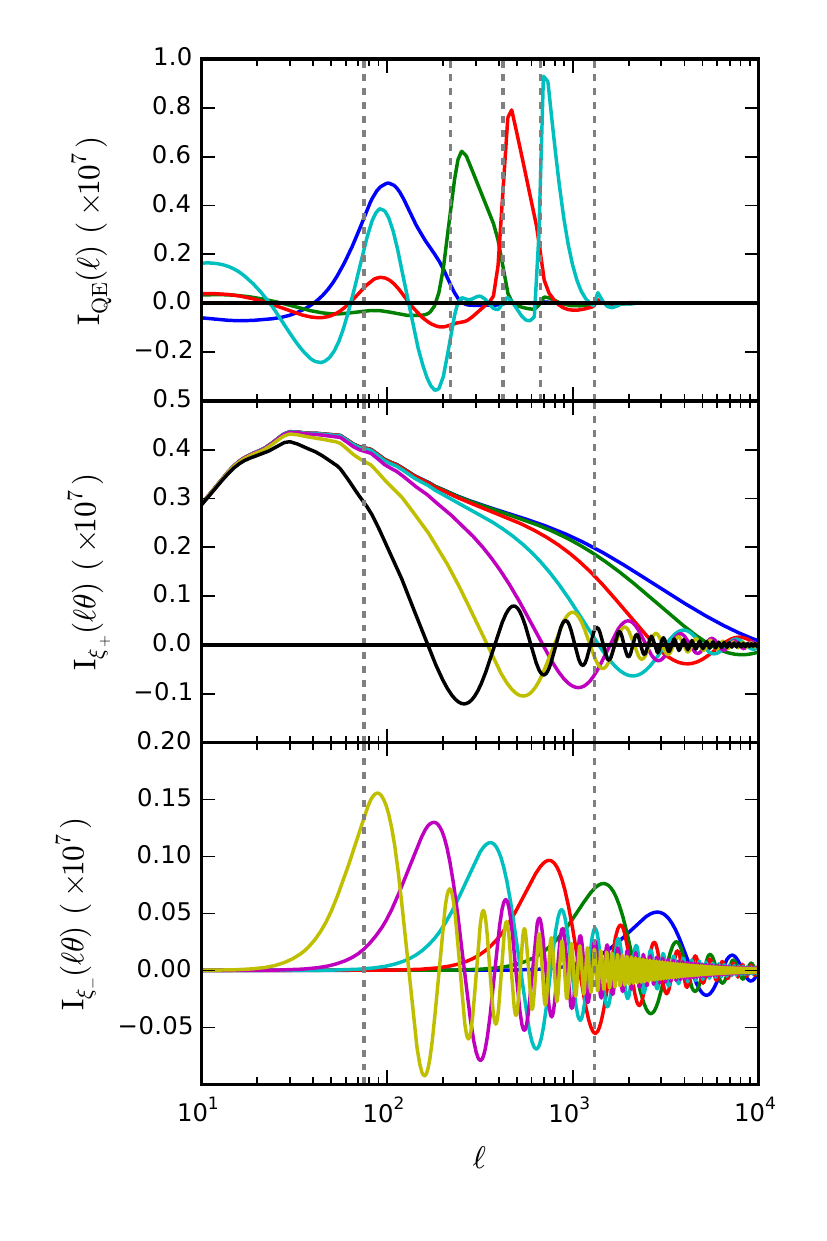}
    \caption{
    Upper panel: measurement of a fiducial shear power spectrum using the quadratic estimator (equation~\ref{eq:qe_comp}) in four band powers between $76 \leq \ell \leq 1310$ and for the lowest redshift bin of the 2 z-bin analysis (Table~\ref{tab:n_eff}). The borders of the bands are indicated by the vertical dashed (grey) lines and each coloured line corresponds to a different band power (Table~\ref{tab:bp_intervals}).
    Mid panel: measurement of the same fiducial shear power spectrum using the $\xi_+$ statistics for correlation functions (equation~\ref{eq:corr_func}) at angular bin centres $\theta$ of 50, 24, 12, 6, 3, 1.5, and $ 0.7 \, {\rm arcmin}$ corresponding to different colours from left to right.
    Lower panel: measurements of the same fiducial shear power spectrum using the $\xi_-$ statistics for correlation functions (equation~\ref{eq:corr_func}) at angular bin centres $\theta$ of 200, 100, 50, 24, 12 and $6 \, {\rm arcmin}$ corresponding to different colours from left to right.}
    \label{fig:theta_comparison}
\end{figure}

A correlation-function based estimator such as the two-point shear correlation function $\xi_{\pm}$ is related to the shear power spectrum $C_{\mu\nu}(\ell)$ at multipoles $\ell$ through
\begin{equation}
\label{eq:corr_func}
\xi^{\mu, \nu}_{\pm}(\theta) = \frac{1}{2 \upi} \int \diff \ell \, \ell C_{\mu\nu}(\ell) J_{0, 4}(\ell \theta) \equiv \int \diff \ell \, I_{\xi_{\pm}}(\ell \theta)\, ,
\end{equation}
where $\theta$ is the angular distance between pairs of galaxies and $J_{0, 4}$ is the zeroth (for $\xi_{+}$) or fourth (for $\xi_{-}$) order Bessel function of the first kind.
In contrast, the quadratic estimator (QE) convolves the shear power spectrum with its band window matrix $W_{A}(\ell)$ (equation~\ref{eq:window}):
\begin{equation}
\label{eq:qe_comp}
\mathcal{B}_A = \sum_{\ell} \frac{\ell(\ell+1)}{2 \upi} W_{A}(\ell) C_A(\ell) \equiv \sum_{\ell} I_{\rm QE}(\ell) \, ,
\end{equation}
where the super-index $A$ runs over all multipole bands and unique redshift correlations.
The convolved power spectra as a function of multipoles defined at the right-hand sides of both equations are shown in Fig.~\ref{fig:theta_comparison} for the lowest redshift bin of the 2 z-bin analysis. In the upper panel we indicate the borders of the bands used in our cosmological analysis (grey dashed lines; see Table~\ref{tab:bp_intervals}). 
In the two lower panels we show the upper and lower limits of our power spectrum analysis. 
For the calculation of $I_{\xi_\pm}(\ell \theta)$ we use the central values of the $\theta_\pm$-intervals from the cosmic shear analysis of \citet{Hildebrandt2016}.
Fig.~\ref{fig:theta_comparison} shows that the $\xi_+$ measurements are highly correlated and anchored at very low multipoles, whereas the $\xi_-$ measurements show a high degree of mode-mixing. In contrast, the quadratic-estimator measurements of the power spectrum are more cleanly separated and the degree of mode mixing is lower. We also note that correlation-function measurements get contributions from lower multipoles than $\ell < 76$ as well as multipoles larger than $\ell > 1310$, which do not contribute to the signal in our power spectrum analysis. At face value most of the cosmological information is contained in high multipoles and although the correlation-function measurements extend further into the high multipole regime, the contributions from these scales are non-negligible only for angular scales $\theta < 3 \, {\rm arcmin}$. 
However, the interpretation of the correlation-function signal at these scales requires accurate knowledge of the non-linear part of the matter power spectrum at high wavenumbers $k$.

Finally, we remark that the disadvantages of the two-point shear correlation functions $\xi_\pm$ described here can also be avoided by using `Complete Orthogonal Sets of E/B-mode Integrals' also known as `COSEBIs' (\citealt{Schneider2010, Asgari2012}; see \citealt{Asgari2017} for an application to data).

\section{Additional figures}
\label{app:extra_fig}

\begin{figure*}
	\centering
	\includegraphics[width=180mm]{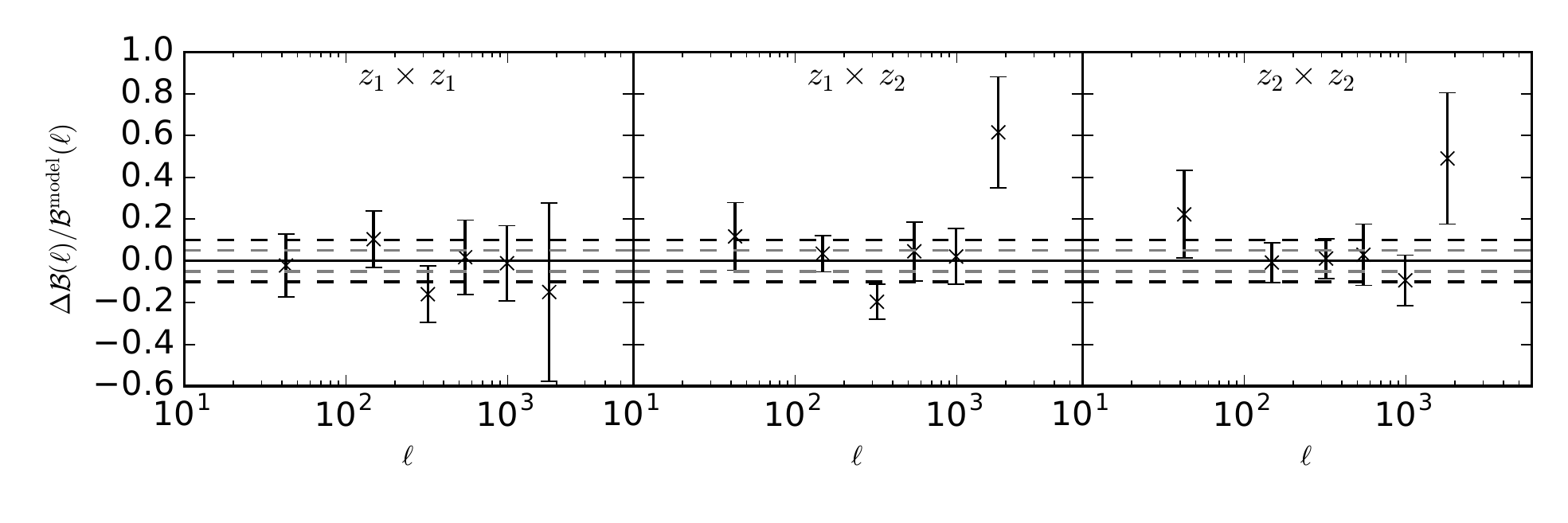}
	\caption{Residuals of input E-mode power and extracted E-modes averaged over 50 CFHTLenS-like GRF realizations of W2. The errors shown here are the errors on the mean band power and hence they are divided by $\sqrt{50}$.}
	\label{fig:comparison_E_modes_GRFs}
\end{figure*}

In Fig.~\ref{fig:comparison_E_modes_GRFs} we show the residuals and the error on the mean between the input and extracted E-mode power from 50 GRF extractions for the highest noise level (Section~\ref{sec:fiducial_B_modes}). The figure shows that the bands considered to enter in the cosmological analysis are unbiased (the first and last band are excluded a priori as discussed in Section~\ref{sec:data_meas}). However, the second-to-last band shows a significant bias (the last band is omitted in the figure because it is off-scale) and therefore it is ignored in the cosmological analysis.

\begin{figure*}
	\centering
	\includegraphics[width=180mm]{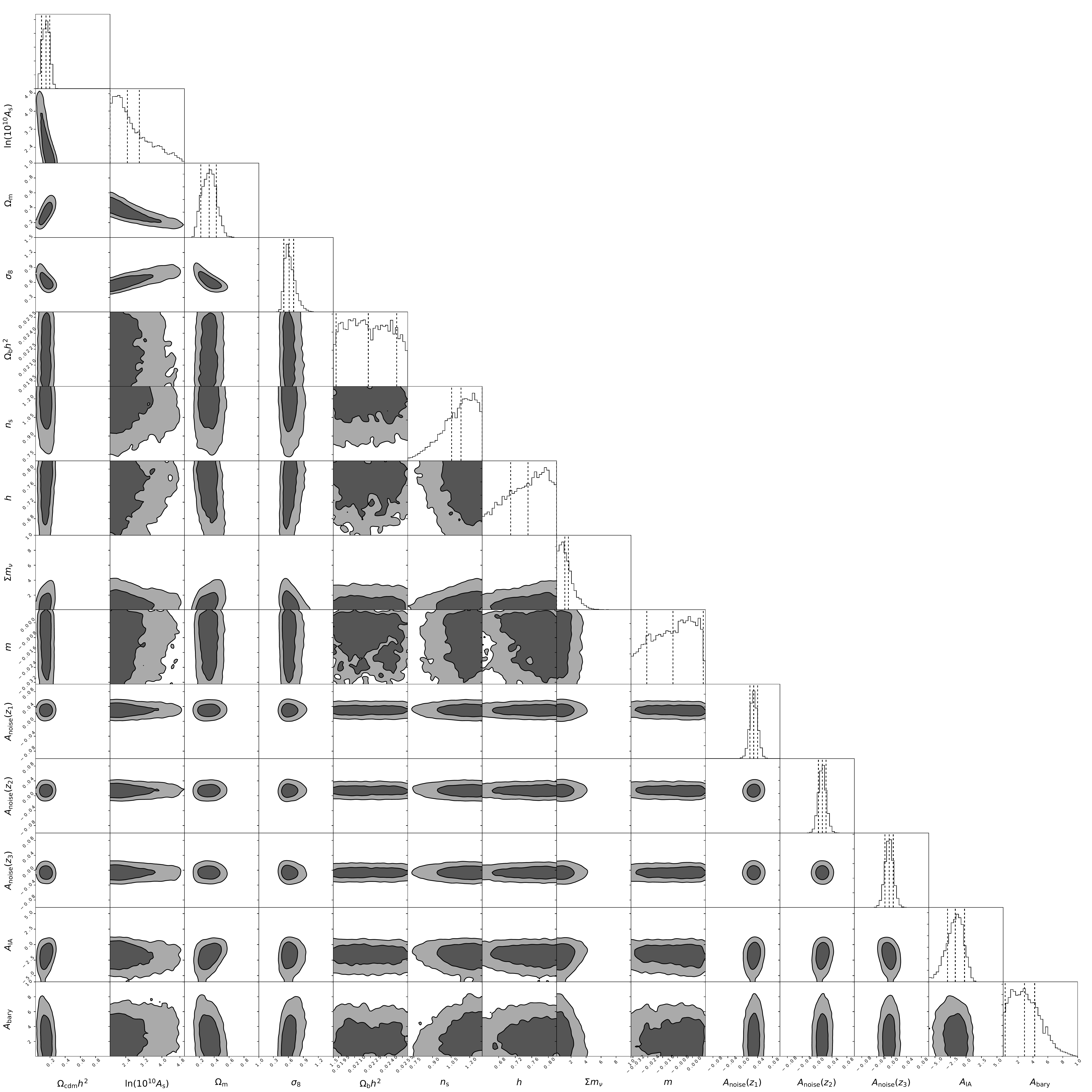} 
	\caption{The parameter constraints derived from sampling the likelihood of the model \lcdm ${+}A_{\rm IA}{+}A_{\rm bary}{+}\Sigma m_\nu{+}{\rm noise}$ for the 3 z-bin analysis (Section~\ref{sec:cosmo_inference} and Table~\ref{tab:cosmo_params}). Note that we also marginalise over the redshift distribution uncertainty and the resetting bias parameters. The parameter $m$ is a dummy variable for the marginalisation over the uncertainties of the multiplicative shear calibration bias (see Section~\ref{sec:m-correction}) and the parameters $A_{\rm noise}(z_\mu)$ describe the amplitudes used in the excess-noise model (see Section~\ref{sec:other_sys}). The dashed lines in the marginalised 1D posteriors denote the weighted median and the 68 per cent credible interval (Table~\ref{tab:cosmo_params}). The contours in each 2D likelihood contour subfigure are $68$ and $95$ per cent credible intervals smoothed with a Gaussian for illustrative purposes only.}
	\label{fig:triangle_base_all}
\end{figure*}

In order to highlight possible parameter degeneracies we show in Fig.~\ref{fig:triangle_base_all} all 2D projections of the parameters used in the most extended model \lcdm $ {+} A_{{\rm IA}} {+} A_{{\rm bary}} {+} \Sigma m_{\nu} {+} {\rm noise}$ (Section~\ref{sec:cosmo_inference} and Table~\ref{tab:cosmo_params}).
 
\begin{figure*}
	\centering
	\begin{subfigure}{0.49\textwidth}
       \centering
       \includegraphics[width=\textwidth]{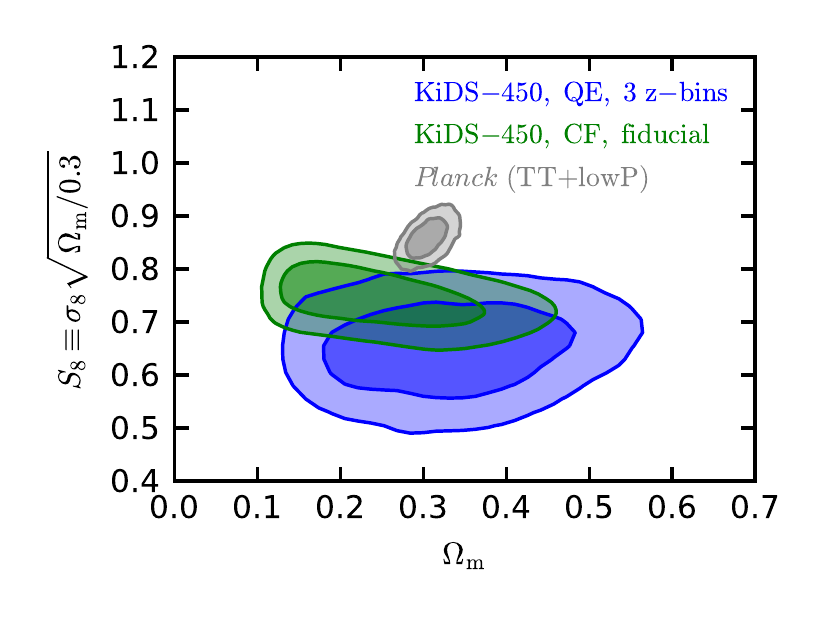}
       \caption{}
    \end{subfigure}
    \begin{subfigure}{0.49\textwidth}
        \centering
        \includegraphics[width=\textwidth]{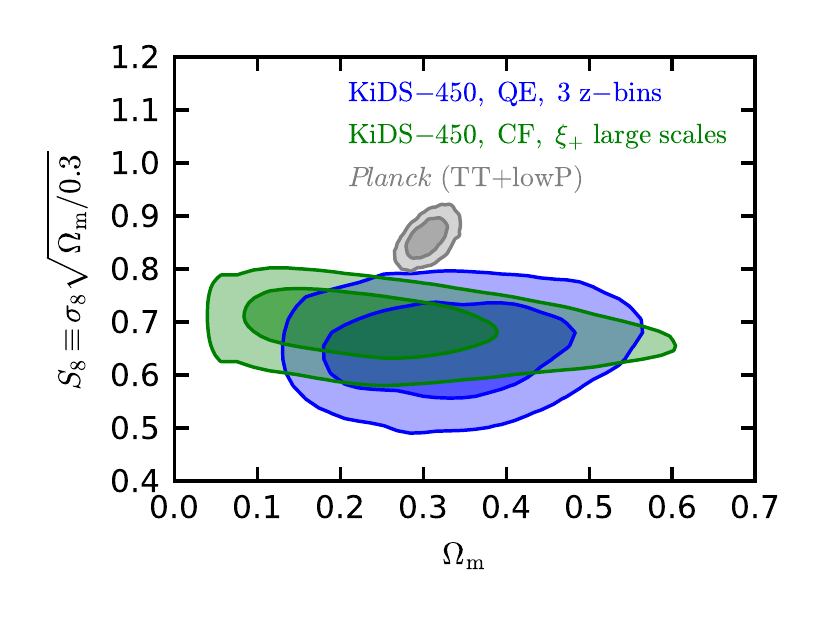}
        \caption{}
    \end{subfigure}
    \caption{(a): Projection of cosmological constraints in the $S_8$ versus $\Omega_{\rm m}$ plane from the KiDS-450 analysis presented here (`KiDS-450, QE, 3 z-bins') and the fiducial correlation-function analysis by \citet[`KiDS-450, CF, fiducial']{Hildebrandt2016}. For comparison we also show contours from \citet[`TT+lowP']{Planck2015_CP}. The inner contours correspond to the 68 per cent credibility interval and the outer ones to the 95 per cent credibility interval. Note that the contours are smoothed with a Gaussian for illustrative purposes only. (b): The same as in (a) but comparing to the `$\xi_{+}$ large scales correlation-function analysis from \citet{Hildebrandt2016}.}
	\label{fig:2d_projection_3zbins}
\end{figure*}

In Fig.~\ref{fig:2d_projection_3zbins} we show constraints in the $S_8$ versus $\Omega_{\rm m}$ plane for the 3 z-bin analysis. 

\section{Sensitivity to large-scale additive bias}
\label{sec:c-term}

Additive biases (equation~\ref{eq:observed_shear}) are mainly caused by a residual PSF ellipticity in the shape of galaxies (e.g. \citealt{Hoekstra2004, vUitert2016}). More generally, any effect causing a preferential alignment of shapes in the galaxy source sample will create an additive bias. For example, in an early stage of the KiDS-450 data processing a small fraction of asteroids ended up in the galaxy source sample. This resulted in strongly aligned shape measurements with very high SNR causing a substantial $c$-term (see appendix~D4 in \citealt{Hildebrandt2016}). This example also demonstrates that a potential $c$-term correction can only be derived empirically from the data.

Here we demonstrate how the quadratic estimator can naturally deal with a residual additive shear in the data. This is a clear advantage over correlation- function statistics that do not separate E- from B-modes such as the $\xi_\pm$ statistics. For these the residual additive shear needs to be properly quantified and subtracted from the data, usually hampering the ability of measuring the cosmic shear signal at large angular separations. This indeed motivated the choice of maximum angular separations used in the cosmological analysis of \citet{Hildebrandt2016}. 

\begin{figure*}
	\includegraphics[width=180mm]{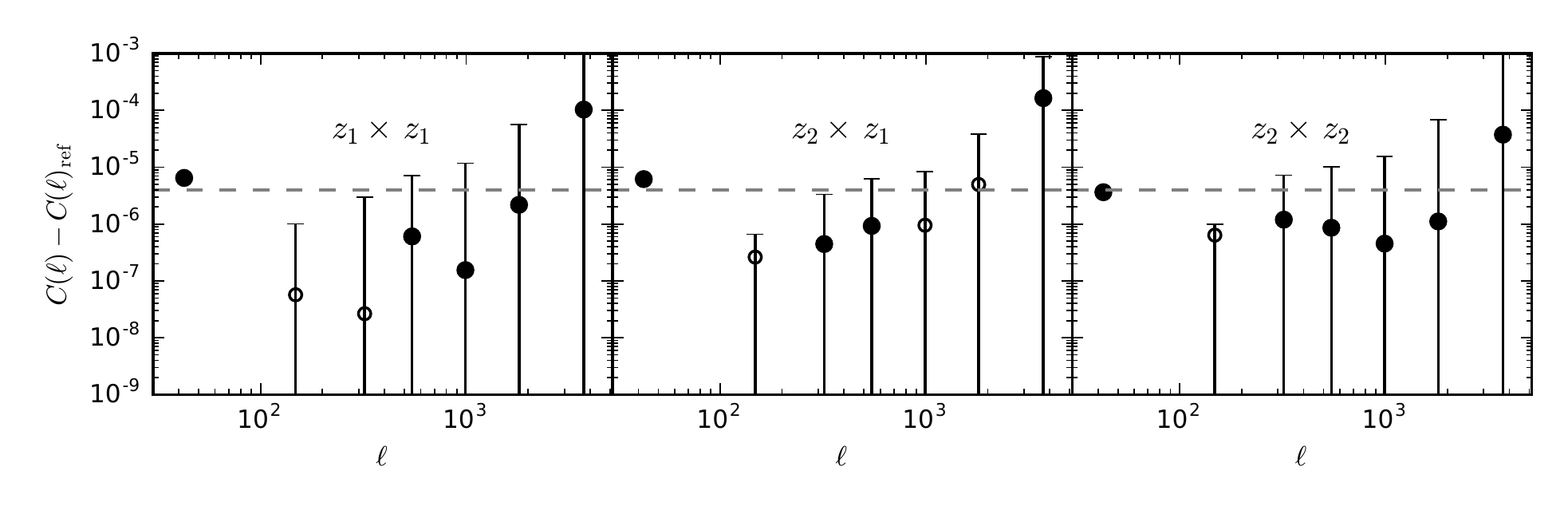}
    \caption{The difference between a shear power spectrum extracted from reference Gaussian random fields (GRFs) and the power spectrum extracted from GRFs in which a global $c$-term of $c = 2 \times 10^{-3}$ was applied to both ellipticity components. From left to right the unique correlations of the two redshift bins are shown. The GRFs were created to match the four fields of CFHTLenS in area, shape, noise properties, and redshift range ($z_1{:} \ 0.50 < z_{\rm B} \leq 0.85$ and $z_2{:} \ 0.85 < z_{\rm B} \leq 1.30$). The signal extraction, however, employs the multipole binning that is also used in the subsequent KiDS data analysis and extends to multipoles significantly below the one set by the field size. The globally applied $c$-term only affects the band power estimate of the first multipole bin but has no effect on the remaining bands. Hence, removing the first band power from a subsequent cosmological analysis is sufficient to account for a leftover global $c$-term in the data. The $1\sigma$ error bars are based on the Fisher matrices and the horizontal dashed (grey) line indicates the square of $c=2 \times 10^{-3}$.}
    \label{fig:ctest}
\end{figure*}

If sufficiently low multipoles are included in the extraction of the first multipole band of the shear power spectrum band powers, this band accounts for any residual DC offset in the data such as the effect of a constant $c$-term. For a clean demonstration of this feature, we employ Gaussian random fields (GRFs) with realistic CFHTLenS survey properties (e.g. masking, noise level; see \citealt{Koehlinger2016} for details). The GRFs were readily available and for this demonstration the differences in survey properties are not of importance. We extract E- and B-modes simultaneously from four GRFs that match the W1, W2, W3, and W4 fields from CFHTLenS in size and shape. The measurements are performed in two broad redshift bins, i.e. $z_1{:} \ 0.50 < z_{\rm B} \leq 0.85$ and $z_2{:} \ 0.85 < z_{\rm B} \leq 1.30$, but we use the same multipole binning as used in the analysis of the KiDS-450 data (Table~\ref{tab:bp_intervals}). For performance reasons we decrease the shear pixel size to $\sigma_{\rm pix}=0\fdg14$. In a first step we extract a reference signal from the GRFs to which no additional global $c$-term was added. In a second step we apply a large but realistic additive term of $c = 2 \times 10^{-3}$ (e.g. fig.~D6 in \citealt{Hildebrandt2016}; \citealt{Jarvis2015}) to both ellipticity components and re-extract the shear power spectra. In Fig.~\ref{fig:ctest} we show the difference between these two signals for all tomographic and multipole bins. As expected, the first multipole bin shows a substantial contamination on the order of the squared global $c$-term, but all remaining bands are essentially unaffected. Hence, removing the first multipole bin from a subsequent cosmological analysis replaces a sophisticated constant $c$-correction at reasonable computational costs. Note, however, that this approach does not account for a spatially varying $c$-correction. However, \citet{Hildebrandt2016} find for the additive bias no significant dependence on the observed position within the field of view for the KiDS-450 tiles. 

\section{Additional tables}
\label{app:extra_tab}

In Tables~\ref{tab:cosmo_params}~and~\ref{tab:combination} we present cosmological parameter constraints for the fiducial model used in the 2 z-bin and 3 z-bin analyses (Section~\ref{sec:cosmo_inference}).

\begin{table*}
\centering
\begin{minipage}{180mm}
	\caption{Cosmological parameter constraints and flat prior ranges.}
	\label{tab:cosmo_params}
	\begin{center}
		\resizebox{\textwidth}{!}{%
		\begin{tabular}{ l r r r r r }
			\toprule 
			Parameters& Flat prior ranges& \lcdm ${+}A_{\rm IA}{+}A_{\rm bary}{+}\Sigma m_\nu{+}{\rm noise}$& \lcdm ${+}A_{\rm IA}{+}A_{\rm bary}{+}\Sigma m_\nu{+}{\rm noise}$\\ 
			& & (2 z-bins)& (3 z-bins)\\			
			\midrule 
			$\Omega_\mathrm{cdm}h^2$& $[0.01, 0.99]$& $0.15_{-0.05}^{+0.04}$& $0.15_{-0.06}^{+0.05}$\\[1mm] 
			$\ln(10^{10}A_\mathrm{s})$& $[1.7, 5.]$& $2.52_{-0.82}^{+0.48}$& $2.47_{-0.77}^{+0.53}$\\[1mm]
			$\Omega_\mathrm{m}$& derived& $0.34_{-0.11}^{+0.09}$& $0.33_{-0.11}^{+0.09}$\\[1mm]
			$\sigma_8$& derived& $0.58_{-0.11}^{+0.09}$& $0.62_{-0.11}^{+0.09}$\\[1mm]
			$\Omega_\mathrm{b}h^2$& $[0.019, 0.026]$& $0.022_{-0.003}^{+0.004}$& $0.022_{-0.003}^{+0.003}$\\[1mm]
			$n_\mathrm{s}$& $[0.7, 1.3]$& $1.08_{-0.13}^{+0.21}$& $1.13_{-0.08}^{+0.17}$\\[1mm]
			$h$& $[0.64, 0.82]$& $0.75_{-0.06}^{+0.07}$& $0.75_{-0.04}^{+0.07}$\\[1mm]
			$\Sigma m_\nu \, ({\rm eV})$& $[0.06, 10.]$& $1.48_{-1.42}^{+0.63}$& $1.16_{-1.09}^{+0.49}$\\[1mm]
			$m$& $[-0.033, 0.007]$& $-0.013_{-0.017}^{+0.017}$& $-0.011_{-0.014}^{+0.016}$\\[1mm]
			$A_{\rm IA}$& $[-6., 6.]$& $-1.81_{-1.21}^{+1.61}$& $-1.72_{-1.25}^{+1.49}$\\[1mm]
			$A_\mathrm{bary}$& $[0., 10.]$& $3.15_{-3.15}^{+1.36}$& $2.87_{-2.60}^{+1.36}$\\[1mm]
			$A_{\rm noise}(z_1)$& $[-0.1, 0.1]$& $0.012_{-0.011}^{+0.011}$& $0.030_{-0.010}^{+0.011}$\\[1mm]
			$A_{\rm noise}(z_2)$& $[-0.1, 0.1]$& $-0.003_{-0.011}^{+0.012}$& $0.014_{-0.011}^{+0.010}$\\[1mm]
			$A_{\rm noise}(z_3)$& $[-0.1, 0.1]$& --& $-0.006_{-0.011}^{+0.011}$\\[1mm]
			$\chi^2$& --& $17.97$& $48.59$\\[1mm]
			dof& --& $12$& $35$\\[1mm]
			\bottomrule 
		\end{tabular}}
	\end{center}
	\medskip 
	\textit{Notes.} We quote weighted median values for each varied parameter (Section~\ref{sec:cosmo_inference}) and derive $1\sigma$-errors using the 68 per cent credible interval of the marginalised posterior distribution. 
\end{minipage}
\end{table*}

\begin{table}
	\caption{Constraints on $S_8$ and $\sigma_8(\Omega_\mathrm{m}/0.3)^\alpha$.}
	\label{tab:combination}
	\begin{center}
		\resizebox{\columnwidth}{!}{%
		\begin{tabular}{ l c c c c }
			\toprule 
			Model& $S_8 \equiv$& Mean error&$\sigma_8$& $\alpha$\\
				 & $\sigma_8\sqrt{\Omega_{\rm m}/0.3}$& on $S_8$& $(\Omega_\mathrm{m}/0.3)^\alpha$& \\			
			\midrule 
			\textbf{2 z-bins:}\\
			$\Lambda \mathrm{CDM}{+}A_{\rm IA}{+}A_{\rm bary}{+}\Sigma m_\nu {+}{\rm noise}$& $0.624_{-0.061}^{+0.069}$& 0.065& $0.623_{-0.062}^{+0.068}$& 0.483\\
			\textbf{3 z-bins:}\\
			$\Lambda \mathrm{CDM}{+}A_{\rm IA}{+}A_{\rm bary}{+}\Sigma m_\nu {+}{\rm noise}$& $0.651_{-0.056}^{+0.060}$& 0.058& $0.650_{-0.056}^{+0.059}$& 0.462\\
			\bottomrule 
		\end{tabular}}
	\end{center}
	\medskip 
	\textit{Notes.} We quote weighted mean values for the constraints on $S_8 \equiv \sigma_8 \sqrt{\Omega_{\rm m}/0.3}$ and $\sigma_8(\Omega_\mathrm{m}/0.3)^\alpha$. The errors denote the 68 per cent credible interval derived from the marginalised posterior distribution.
\end{table}


\bsp	
\label{lastpage}
\end{document}